\documentclass[aps,pra,twocolumn,superscriptaddress]{revtex4-2}
\usepackage{amsfonts}
\usepackage{graphicx}
\usepackage{epstopdf}
\usepackage{epsfig}
\usepackage{amsmath}
\usepackage[normalem]{ulem}
\usepackage{multirow}
\usepackage{makecell}
\usepackage{array}
\usepackage{booktabs}
\usepackage{mathtools}
\usepackage{bm}
\usepackage{color}
\usepackage{soul,xcolor}
\usepackage{array}
\usepackage{booktabs}
\usepackage{hyperref}

\usepackage{tabularx}
\newcolumntype{Y}{>{\centering\arraybackslash}X}

\usepackage{blindtext}
\usepackage{comment}
\definecolor{lightblue}{rgb}{0.3, 0.3, 0.90}
\definecolor{darkblue}{rgb}{0.0, 0.2, 0.5}
\definecolor{lightred}{rgb}{1.0, 0.6, 0.6}
\definecolor{darkred}{rgb}{0.8, 0.0, 0.0}

\begin{document}
\title{Controlling nonergodicity in quantum many-body systems by reinforcement learning}

\author{Li-Li Ye}
\affiliation{School of Electrical, Computer and Energy Engineering, Arizona State University, Tempe, Arizona 85287, USA}

\author{Ying-Cheng Lai} \email{Ying-Cheng.Lai@asu.edu}
\affiliation{School of Electrical, Computer and Energy Engineering, Arizona State University, Tempe, Arizona 85287, USA}
\affiliation{Department of Physics, Arizona State University, Tempe, Arizona 85287, USA}

\date{\today}
\begin{abstract}

Finding optimal control strategies to suppress quantum thermalization for arbitrarily initial states, the so-called quantum nonergodicity control, is important for quantum information science and technologies. Previous control methods relied largely on theoretical model of the target quantum system, but invertible model approximations and inaccuracies can lead to control failures. We develop a model-free and deep reinforcement-learning (DRL) framework for quantum nonergodicity control. It is a machine-learning method with the unique focus on balancing exploration and exploitation strategies to maximize the cumulative rewards so as to preserve the initial memory in the time-dependent nonergodic metrics over a long stretch of time. We use the paradigmatic one-dimensional tilted Fermi-Hubbard system to demonstrate that the DRL agent can efficiently learn the quantum many-body system solely through the interactions with the environment. The optimal policy obtained by the DRL provides broader control scenarios for managing nonergodicity in the phase diagram as compared to, e.g., the specific protocol for Wannier-Stark localization. The continuous control protocols and observations are experimentally feasible. The model-free nature of DRL and its versatile search space for control functions render promising nonergodicity control in more complex quantum many-body systems.

\end{abstract}
\date{\today}

\maketitle

\section{Introduction} \label{sec:intro}

Quantum nonergodicity has been recognized as a central concept in out-of-equilibrium 
quantum dynamical systems~\cite{anderson:1958,emin:1987,abanin:2019,vu:2022,kohlert:2019}. 
Relevant physical phenomena include spatial localization such as 
Anderson~\cite{anderson:1958} and Wannier-Stark localization~\cite{emin:1987,bhakuni:2019}, and Hilbert space localization such as quantum many-body scars (QMBS)~\cite{serbyn:2021} and many-body 
localization (MBL)~\cite{abanin:2019}. The unique attribute of quantum nonergodicity 
in suppressing thermalization has implications to fields ranging from statistical 
mechanics~\cite{nandkishore:2015} to quantum information science and 
technologies~\cite{parameswaran:2018}. Nonergodicity in quantum many-body systems can be 
generated by a number of physical mechanisms, each leading to rich and complex quantum 
phases. Because of the potential of broad applications, controlled generation of quantum
nonergodicity has attracted a great deal of recent attention~\cite{adler:2024,kohlert:2023,desaules:2021,huang:2024strongly,scherg:2021,hudomal:2022,su:2023,bluvstein:2021,zhao:2020,mizuta:2020,sugiura:2021,huang:2024,po:2016,sonner:2021,ponte:2015,geraedts:2016,sierant:2023}. 

Quantum nonergodicity control aims to find optimal control protocols to suppress quantum 
ergodicity for diverse initial states. For example, in Floquet engineering, controlled 
realization of QMBS and MBL through periodic driving is experimentally feasible, but the 
space of control functions is often limited due to its periodic nature, such as sinusoidal 
driving~\cite{hudomal:2022,su:2023,bluvstein:2021,zhao:2020}, periodic pulse 
control~\cite{maskara:2021,mukherjee:2020}, square wave~\cite{diringer:2021} and binary 
driving control of two distinct non-commuting Hamiltonians applied in 
sequence~\cite{mizuta:2020,sugiura:2021,huang:2024,po:2016,sonner:2021,ponte:2015,geraedts:2016,sierant:2023}, 
as well as Floquet automata circuit control~\cite{rozon:2022}. In Floquet engineering,
the optimization of the periodic control has generally been based on theoretical 
models~\cite{chandran:2023}, where control is optimized by the prior knowledge about the 
physics of the target system, such as the intrinsic dynamics of the quantum many-body 
system including subharmonic response and discrete 
time-cystalline~\cite{maskara:2021,bluvstein:2021} as well as conservation and symmetry 
properties~\cite{huang:2024,haldar:2021,pai:2019} of such systems. Alternatively, adiabatic 
approaches and their extensions such as counterdiabatic 
driving~\cite{demirplak:2003,petiziol:2018,sels:2017} and quantum leakage 
minimization~\cite{ljubotina:2022,ljubotina:2024} combine analytical and numerical methods 
to find the optimal protocol along a trajectory in the parameter space. These methods 
provide physical insights and interpretation but inevitably suffer from model inaccuracies
and approximations. It is worth noting that traditional optimal control methods such as 
gradient-free optimal control (e.g., chopped random 
basis~\cite{caneva:2011,doria:2011,muller:2022}) and gradient-based optimal control 
(e.g., gradient ascent pulse 
engineering~\cite{jensen:2021,jensen:2021achieve,machnes:2011,agundez:2017}) were also 
model-based. While there were works on controlling quantum systems to a specific target 
state~\cite{Bukov:2018,metz:2023} or in entanglement engineering~\cite{ye:2024}, to our 
knowledge, model-free approaches have not been investigated for nonergodicity control in
quantum many-body systems.

In this paper, we develop a deep reinforcement-learning (DRL) based framework for quantum 
nonergodicity control. In machine-learning based control, reinforcement learning (RL) has 
emerged as an effective model-free approach with the capability of finding the optimal 
strategy in a vast and versatile search space of control functions and the ability to 
adaptively discover control strategies that the traditional methods tend to 
overlook~\cite{kaelbling:1996,wang:2020deep}. RL employs a trial-and-error learning 
process to maximize the cumulative rewards through exploration and exploitation in search 
for a globally optimal policy. DRL further enhances these capabilities by using deep neural 
networks to optimize the RL agent. In quantum systems, model-free RL control is capable of 
generating policies or value functions based solely on the interactions with the quantum 
environment, without any prior knowledge of the model of that 
environment~\cite{ramirez:2022}, in contrast to model-based DRL methods~\cite{kaiser:2019} 
that employ a pre-built model of the environment to guide policy decisions. One issue is
choosing an algorithm that is particularly appropriate for quantum nonergodicity control.
We choose the proximal policy optimization (PPO) 
algorithm~\cite{schulman:2017,wang:2020truly}, which is justified, as follows.

In modern machine learning, a number of DRL algorithms have been developed, including 
those incorporating the trust region proximal optimization (TRPO) algorithm~\cite{wu:2017}, 
deep-Q network (DQN)~\cite{mnih:2015}, and deep deterministic policy gradient 
(DDPG)~\cite{lillicrap:2015}. Specifically, TRPO provides a common solution to the local 
minima challenge in optimal policy search. By confining policy updates within a trust 
region, TRPO ensures that policies do not deviate too far from previous policies. This 
mechanism not only enhances stability during the training of the RL agent but also 
ensures a monotonic improvement in policy search. Among the DQN and DDPG algorithms, PPO 
stands out as a state-of-the-art algorithm~\cite{schulman:2017,wang:2020truly}. As a hybrid 
actor-critic approach, PPO adeptly navigates the delicate balance between reducing the 
variance of policy gradients and diminishing bias linked to the value functions. Operating 
within the framework of deep neural networks, PPO effectively addresses the curse of 
dimensionality, thereby enhancing its applicability and scalability in complex environments.
PPO achieves data efficiency and reliable performance of TRPO but with a first-order 
optimization procedure, facilitating implementation with reduced computational complexity.

To demonstrate model-free DRL to achieve nonergodicity control in quantum many-body 
systems in a concrete setting, we employ the 1D tilted Fermi-Hubbard model, a paradigm in 
the study of quantum many-body systems capable of generating a spectrum of weak 
ergodicity-breaking phenomena~\cite{scherg:2021,desaules:2021,kohlert:2023}. Our PPO 
agent relies exclusively on the observables and rewards it receives at each step to make 
decisions, without requiring an explicit physical model for policy decisions. There are 
two possible physical observables. The first is spin-resolved imbalance~\cite{scherg:2021}, 
the normalized differences in the occupation numbers of spin-up and spin-down particles 
between odd and even lattice sites. The sceond is fidelity that measures the square of 
the norm of the overlap between the time-evolved quantum state and the initial quantum 
state based on the full chain, partial chain, or even just a single 
site~\cite{guo:2021,reiner:2016,zhang:2023}. A common attribute of quantum nonergodic 
quantities is their retention of the initial memory over the long time, so the reward 
function at each time step is directly linked to the observation and can be maximized 
when the time-evolved observation is consistent with its initial value. The policy is 
optimized by maximizing the accumulated reward. From this perspective, the DRL agent 
focuses solely on minimizing the discrepancy between the accumulated fidelity or 
imbalance and its initial value. As a result, the agent discovers the optimal policy 
through direct engagement with the quantum environment, a hallmark of model-free DRL, 
making it possible for the control algorithm to be implemented in experiments. Another 
distinct feature of our work, which facilitates experimental implementation, is the use of 
partial observations or even just a single site observation to control nonergodicity in 
many-body quantum systems, in contrast to previous works~\cite{Bukov:2018,metz:2023} in 
this field that required the complete observations of the current quantum state. 

In Sec.~\ref{subsec:model}, we introduce the 1D tilted Fermi-Hubbard model and outline the 
experimental setting for the observation and control space. In Sec.~\ref{subsec:PPO}, we 
clarify the basic concept of PPO agent for quantum nonergodicity control. In 
Sec.~\ref{sec:results}, we demonstrate the performance of DRL with partial observations. 
Section~\ref{sec:physics} provides a physical interpretation and understanding of the 
optimal policy derived from DRL. Conclusions and a discussion about the limitations 
and potential future research are offered in Sec.~\ref{sec:discussion}. 

\section{Model and control method} \label{sec:model_method}

\subsection{One-dimensional tilted Fermi-Hubbard model} \label{subsec:model}

The Fermi-Hubbard or Bose-Hubbard model with a tilted potential has garnered a great deal 
of recent interest due to the emerging new physics that may arise commonly in many other
quantum many-body systems~\cite{guardado:2020,zechmann:2022,zhang:2020,oppong:2022,kiely:2024,adler:2024,kohlert:2023,desaules:2021,huang:2024strongly,scherg:2021,boesl:2024,lake:2023,morong:2021,guo:2021,guo:2021Stark}. 
For example, the tilted potential breaks the translational invariance and integrability, 
and so can induce subdiffusive transport due to its coupling to mass transport in 
mass-imbalanced 1D or 2D tilted Fermi-Hubbard 
models~\cite{guardado:2020,zechmann:2022,zhang:2020,oppong:2022,kiely:2024}. The 
subdiffusive property is related to the nature of nonergodic dynamics~\cite{guardado:2020}. 
Moreover, the tilted potential can lead to phenomena such as Hilbert space 
fragmentation~\cite{adler:2024,kohlert:2023}, QMBS~\cite{desaules:2021,huang:2024strongly}, 
 and deconfinement dynamics of 
fractons~\cite{boesl:2024} and non-Fermi liquids~\cite{lake:2023}. The tilted potential 
model finds applications across diverse systems such as ultracold fermions in tilted 
optical lattices~\cite{scherg:2021,kohlert:2023}, trapped ions~\cite{morong:2021}, and 
superconducting qubits~\cite{guo:2021}. Motivated by experimental breakthroughs and the 
nonergodicity induced by a titled potential, we use the 1D tilted Fermi-Hubbard model to 
investigate model-free nonergodicity control. 

The Hamiltonian of the 1D tilted Fermi-Hubbard chain model is~\cite{desaules:2021} 
\begin{align} \nonumber
	\hat{H} &= \sum\limits_{j,\sigma=\uparrow(\downarrow)}\left(-J\hat{c}^{\dagger}_{j+1,\sigma}\hat{c}_{j,\sigma}+\textnormal{h.c.}+\Delta j\hat{n}_{j,\sigma}\right) \\ \label{eq:main_eq}
	&+ U\sum\limits_{j}\hat{n}_{j,\uparrow}\hat{n}_{j,\downarrow},
\end{align}
where $J$ denotes the nearest-neighbor coupling, $\Delta$ is a uniform tilted 
potential distributed in position space, and $U$ is the on-site Coulomb interaction. 
The Hamiltonian includes the fermionic creation ($\hat{c}^{\dagger}_{j,\sigma}$) and 
annihilation ($\hat{c}_{j,\sigma}$) operators, as well as the number operator 
$\hat{n}_{j,\sigma}=\hat{c}^{\dagger}_{j,\sigma}\hat{c}_{j,\sigma}$. For simplicity, 
we consider a lattice with $\mathcal{N}$ sites, where the spin-up and spin-down fermions
are equally distributed, denoted by
\begin{align} \nonumber
\mathcal{N}_{\uparrow}=\mathcal{N}_{\downarrow}=\mathcal{N}/2. 
\end{align}
This setup implies an electron filling factor of $\nu=1$, as in a previous work in the
Fock basis~\cite{desaules:2021}. We assume periodic boundary conditions, where the spin 
direction is maintained when a particle hops crosses the boundary. At any given lattice 
site, the occupation by the spin-up or spin-down electrons is represented as 
$|\uparrow\rangle$ or $|\downarrow\rangle$, respectively, while an empty site is indicated 
as $|0\rangle$. A site simultaneously occupied by both spin up and down electrons, known 
as a doublon~\cite{desaules:2021}, is denoted by $|\updownarrow\rangle$.

Experimentally, a number of platforms are available for controlling the system described 
by the 1D tilted Fermi-Hubbard chain model. For example, in an optical lattice, a tilted 
potential can be modulated by a magnetic field gradient~\cite{scherg:2021,kohlert:2023} and 
the on-site Coulomb interaction $U$ is tunable via a magnetic Feshbach 
resonance~\cite{scherg:2021,schreiber:2015}. Specifically, the tilted Fermi-Hubbard model 
in an optical lattice can be characterized by the imbalance~\cite{scherg:2021}:
\begin{equation}
\mathcal{I}^{\uparrow(\downarrow)}\equiv\frac{\mathcal{N}^{\uparrow(\downarrow)}_{\textnormal{o}}-\mathcal{N}^{\uparrow(\downarrow)}_{\textnormal{e}}}{\mathcal{N}^{\uparrow(\downarrow)}_{\textnormal{o}}+\mathcal{N}^{\uparrow(\downarrow)}_{\textnormal{e}}},
\end{equation}
where $\mathcal{N}^{\uparrow(\downarrow)}_{\textnormal{o}}$ and 
$\mathcal{N}^{\uparrow(\downarrow)}_{\textnormal{e}}$ are the occupation numbers of the 
spin-up and spin-down electrons at the odd and even lattice sites, respectively. There are 
also experimental techniques~\cite{gimperlein:2005,greschner:2014,schreiber:2015,sierant:2018,zhao:2020} that allow for an independent manipulation of $\Delta(t)$ and $U(t)$ over time, 
offering precise control over the system's dynamics. In addition, superconducting qubits 
quantum simulators~\cite{guo:2021,reiner:2016,zhang:2023} with an integrated and 
programmable large-scale platform offer flexibility in the control protocols, where 
quantum tomography measurements offer direct experimental access to the components of the 
reduced density matrix~\cite{guo:2021,reiner:2016,zhang:2023}. This capability enables 
precise measurements of the fidelity~\cite{guo:2021,zhang:2023} for both the sub-chain
and the full chain, denoted as $\mathcal{F}_{\textnormal{sub}}$ and 
$\mathcal{F}_{\textnormal{full}}$, respectively. In fact, the time evolution of Von Neumann 
entanglement entropy for the sub-chain has been observed~\cite{zhang:2023}. The 
computational approaches for evaluating the nonergodic metrics are described in 
Appendix~\ref{appendix_A}.

For convenience and clarity, in the following, we present our results using the time unit 
$\tau \equiv \hbar/J$ with the reduced Planck constant $\hbar$ and the nearest-neighbor 
coupling strength $J$. The potential terms $\Delta$ and $U$ are expressed in units of $J$, 
as outlined in Appendix~\ref{appendix_A}. The quantum dynamics of the 1D tilted 
Fermi-Hubbard chain are governed by the Schr\"{o}dinger equation. We employ Trotter
decomposition~\cite{trotter:1959,suzuki:1976} with the discrete time step of size 
$dt=0.005\tau$. A detailed accuracy analysis can be found in Appendix~\ref{appendix_A}.

\subsection{Method of DRL based quantum nonergodicity control} \label{subsec:PPO}

\begin{figure} [h!]
\centering
\includegraphics[width=\linewidth]{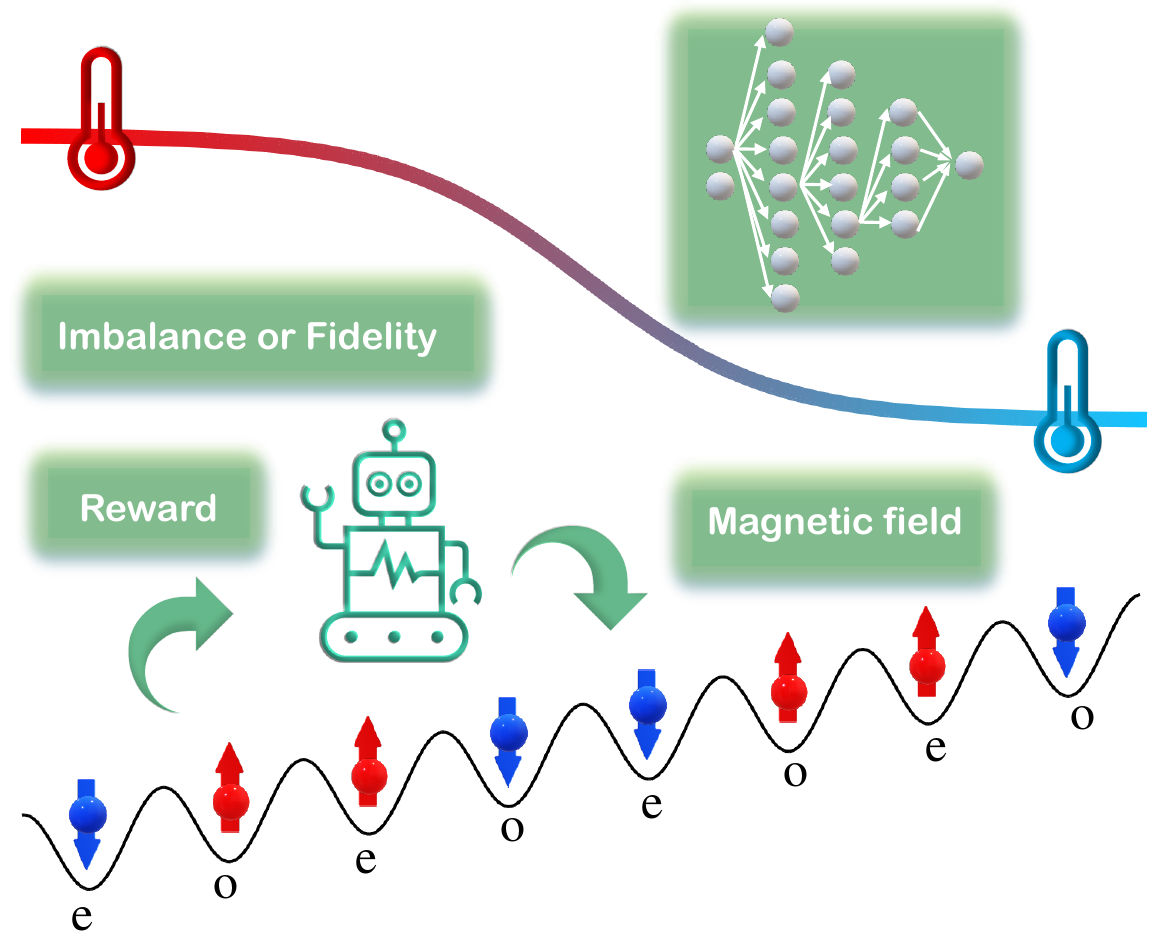}
\caption{Scenario of quantum nonergodicity control by model-free DRL. The PPO algorithm 
coupled with neural networks is used for control. The task involves training a randomly 
initialized agent to discover the optimal control protocol for steering quantum 
nonergodicity in the 1D tilted Fermi-Hubbard model. This system is initialized in a 
specific quantum state within the Fock space. The observation space is one of the 
following: (1) imbalance metrics $[\mathcal{I}^{\uparrow},\mathcal{I}^{\downarrow}]$ 
representing the normalized differences in the occupation numbers of spin-up and spin-down 
particles between the odd and even lattice sites, (2) partial fidelity 
$\mathcal{F}_{\textnormal{sub}}$ that provides a partial view of the lattice chain 
dynamics, and (3) full fidelity $\mathcal{F}_{\textnormal{full}}$ that offers a 
complete observation of the entire lattice chain. The agent receives the corresponding 
rewards, with the maximum value specifically designed to incentivize the maintenance of 
the initial state throughout the time evolution. Following policy updates aimed at 
maximizing the cumulative reward over time, the agent intervenes in the quantum many-body system by applying a designated tilted potential and on-site Hubbard potential, which can be experimentally implemented via a physical control field, such as a magnetic field. As the accumulated reward converges, the 
degree of quantum ergodicity gradually diminishes.}
\label{fig:scheme_RL_QMBS}
\end{figure}

Quantum nonergodicity describes out-of-equilibrium phenomenon that arise when a quantum 
system resists thermalization or equilibration even after long time evolution. It implies 
that time-evolved quantum states retain the memory of their initial conditions, remaining 
closely aligned with them over prolonged duration. Training a DRL agent to realize quantum 
nonergodicity control thus entails maintaining the time-evolved quantum states as closely 
as possible to their initial, pure, and unentangled states throughout the evolution 
process. The scenario of DRL training is illustrated in Fig.~\ref{fig:scheme_RL_QMBS}. 

The DRL training contains the following components.

\paragraph*{Initialization.} The environment for the DRL PPO agent to learn is the quantum 
many-body system: the 1D tilted Fermi-Hubbard chain formulated in Eq.~\eqref{eq:main_eq}. The 
quantum state is initialized in the Fock space. For example, for a lattice of size 
$\mathcal{N}=8$, two initial states are
\begin{align} \nonumber
	|-+-+\rangle &=|\downarrow\uparrow\uparrow\downarrow\downarrow\uparrow\uparrow\downarrow\rangle, \\ \nonumber
	|+-+-\rangle &=|\uparrow\downarrow\downarrow\uparrow\uparrow\downarrow\downarrow\uparrow\rangle, 
\end{align}
where $|-\rangle \equiv |\downarrow\uparrow\rangle$ and 
$|+\rangle \equiv |\uparrow\downarrow\rangle$. For conciseness, we denote the two kinds of
initial states as $|-+\rangle$ and $|+-\rangle$, respectively. In a lattice system
with open boundaries, under the approximation $\Delta\approx U\gg J$ these states are 
in fact QMBS states in the corresponding effective Hamiltonian model~\cite{desaules:2021}.
Other permutations of spin configurations in the Fock space tested in our work include 
$|\uparrow\downarrow\uparrow\downarrow\uparrow\downarrow\uparrow\downarrow\rangle$, $|\uparrow\downarrow\uparrow\uparrow\downarrow\downarrow\uparrow\downarrow\rangle$, $|\downarrow\uparrow\downarrow\uparrow\downarrow\uparrow\downarrow\uparrow\rangle$, $|\uparrow\downarrow\uparrow\downarrow\uparrow\uparrow\downarrow\downarrow\rangle$, and $|\uparrow\uparrow\downarrow\downarrow\uparrow\uparrow\downarrow\downarrow\rangle$.

\paragraph*{Observation Space.} Three physical quantities are used for the DRL agent to 
observe the environment: (1) the imbalance vector 
$[\mathcal{I}^{\uparrow}(t),\mathcal{I}^{\downarrow}(t)]$ with 
$\mathcal{I}^{\uparrow(\downarrow)}(t)\in[-1,1]$, (2) partial-chain fidelity 
$\mathcal{F}_{\textnormal{sub}}(t)\in[0,1]$, and (3) full-chain fidelity 
$\mathcal{F}_{\textnormal{full}}(t)\in[0,1]$. It is worth noting that, in experiments, 
detecting imbalance or partial fidelity, or even simply observing a single lattice site, 
can be more efficient than observing the full-chain fidelity. However, having only partial 
information about the quantum system poses challenges for optimizing control protocols.

\paragraph*{Action Space.} The configuration of the observation and action spaces 
determines computational and control complexity. We adopt continuous observation and action 
spaces for optimal policy search. Specifically, the action space comprises the tilted potential $\Delta(t)$ and the on-site Hubbard term
$U(t)$, each ranging from $-10J$ to $10J$. The limited range of these global control fields is chosen to avoid trivial solutions, such as ideal Wannier-Stark localization ($\Delta\gg J$, and $U=0$), which would lead to nonergodic behavior but are restricted by experimental relevance. Despite the discretization of time evolution, the values at each time step remain continuous within the range $[-10, 10] J$, and can be parameterized by a deep neural network. This defines a continuous action space with infinite dimensions.

\paragraph*{Reward Design.} The design of reward functions is tailored to the quantity of 
observation. For the imbalance vector, the reward function takes the form:
\begin{equation}
    R(t) = -|\mathcal{I}^{\downarrow}(t)-\mathcal{I}^{\downarrow}(0)| -|\mathcal{I}^{\uparrow}(t)-\mathcal{I}^{\uparrow}(0)|.
\end{equation}
Alternatively, if the observation space involves the sub- or full-chain fidelity, the 
reward function becomes:
\begin{equation}
    R(t)=-|\sqrt{\mathcal{F}(t)}-1|.
\end{equation}
Under this setup, the agent incurs a negative penalty for deviations from the initial 
states, encouraging it to maintain proximity to the initial configuration.

\paragraph*{Training.} The training of the DRL agent relies on a delicate balance of 
exploration and exploitation strategies, which is crucial for learning the optimal policy 
to maintain quantum nonergodicity. In particular, achieving this balance is essential for 
uncovering the effective strategies for sustaining quantum nonergodicity. The DRL agent 
must explore diverse actions to comprehend their impact on the quantum system, while also 
exploiting established strategies to maximize the reward. This iterative process requires 
that the agent interact with the quantum system, observe the resulting states, and improve 
its policy based on the received rewards, as shown in Fig.~\ref{fig:scheme_RL_QMBS}. As 
outlined in Appendix~\ref{appendix_B}, both the actor and critic utilize the independent 
neural networks with the identical size. For various tasks, we adopt two alternative 
neural-network configurations with three hidden layers in a multilayer perceptron: 
{\bf Config A} - $\textnormal{NN}=[256, 128, 64]$ with a learning rate of $10^{-4}$; 
{\bf Config B} - a smaller neural network $\textnormal{NN}=[128, 64, 32]$ with the learning 
rate $0.5\times 10^{-3}$. The PPO agent is implemented using the Reinforcement Learning 
Toolbox in MATLAB.

\section{DRL based quantum nonergodicity control: Results} \label{sec:results}

\subsection{Illustration of control performance}

\begin{figure*} [ht!]
\centering
\includegraphics[width=\linewidth]{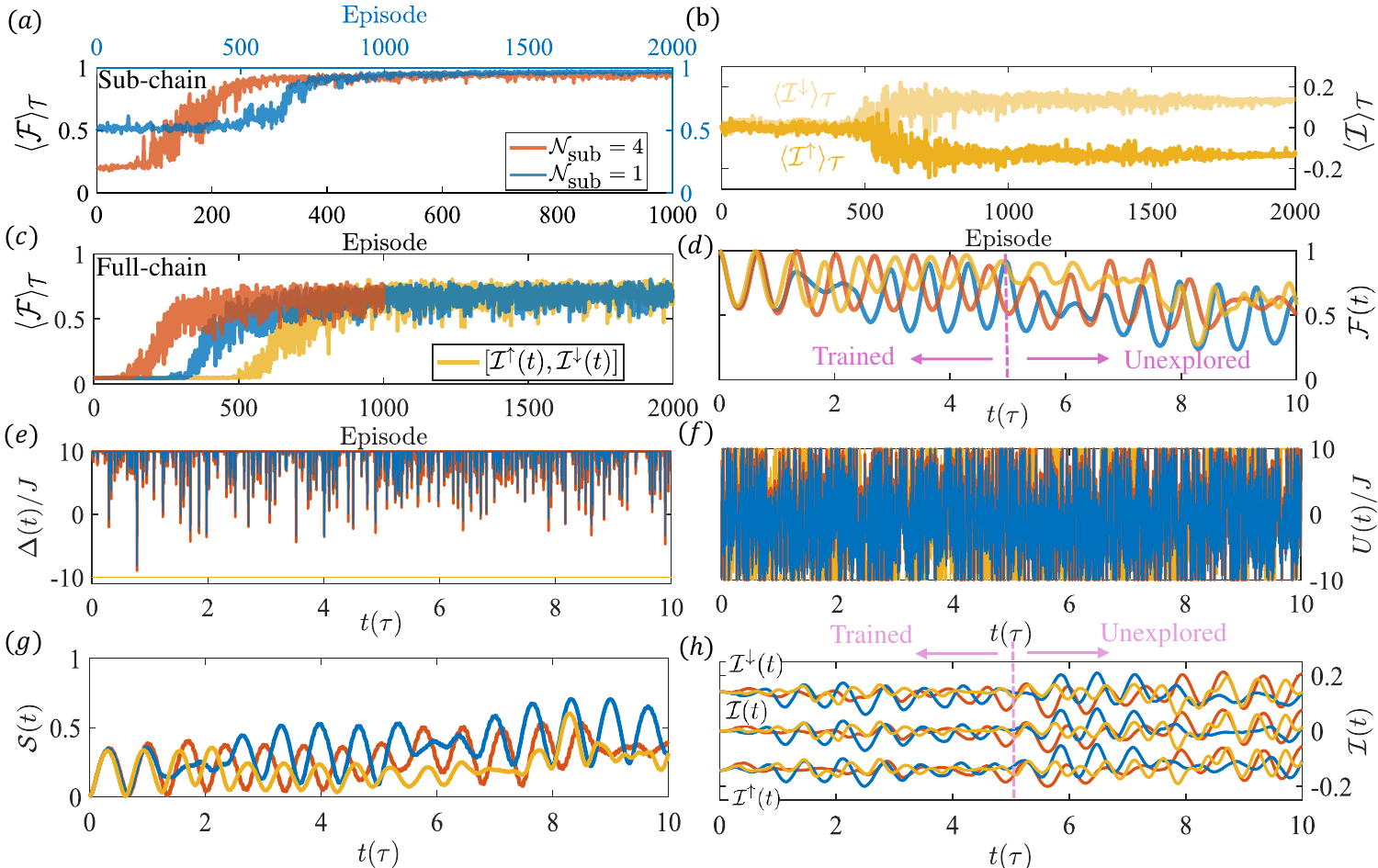}
\caption{Evaluation of quantum nonergodicity control by DRL agent. The 1D tilted 
Fermi-Hubbard chain has the size $\mathcal{N}=14$. The observations are performed using 
(1) sub-chain fidelity $\mathcal{F}_{\textnormal{sub}}$ with 
$\mathcal{N}_{\textnormal{sub}}=1$ or $4$ sites from the left-hand side of the chain, 
and (2) the imbalance vector $[\mathcal{I}^{\uparrow},\mathcal{I}^{\downarrow}]$ from the 
entire chain. These options constitute distinct tasks shown with different colors in (a-h), 
with $\mathcal{N}_{\textnormal{sub}}=1,\;4$ by $\mathcal{F}_{\textnormal{sub}}$, and the 
$[\mathcal{I}^{\uparrow},\mathcal{I}^{\downarrow}]$ tasks in blue, orange, and yellow 
curves, respectively. The initial state for both tasks is identical: $|-+-+-+-\rangle$. 
(a,b) Training phase for the $\mathcal{F}_{\textnormal{sub}}$ and 
$[\mathcal{I}^{\uparrow},\mathcal{I}^{\downarrow}]$ tasks, respectively. The learning 
curves chart the average fidelity $\langle \mathcal{F}\rangle_{\mathcal{T}}$ and the 
average imbalance $\langle \mathcal{I}\rangle_{\mathcal{T}}$ over the time horizon 
$\mathcal{T}=5\tau$ in episodes, reflecting episodic learning with policy updates and 
quantum state resets after each episode. (c) Convergence of the resulting 
$\langle \mathcal{F}\rangle_{\mathcal{T}}$ within the full chain, which agrees with that 
of the sub-chain task in (a) and the imbalance task in (b). (d-h) Results from the testing 
phase, where the test time horizon is $\mathcal{T}=10\tau$, encompassing the unexplored 
capabilities of the DRL agent. The nonergodic metrics, including 
$\mathcal{F}(t), \mathcal{S}(t)$ and $\mathcal{I}(t)$, demonstrate the success of 
nonergodicity control through the optimal action flow discovered by the DRL agent, as 
depicted in (e) and (f). The neural network size for the two tasks with 
$\mathcal{N}_{\textnormal{sub}}=1$ and $\mathcal{I}^{\uparrow(\downarrow)}$ is from 
{\bf Config A}, and that with $\mathcal{N}_{\textnormal{sub}}=4$ is from {\bf Config B} 
(specified in Sec.~\ref{subsec:PPO}).}
\label{fig:numLattice_14_T_5_F_sub_imbalance}
\end{figure*}

The distinct feature of quantum nonergodicity lies in its capacity to preserve the memory 
of the initial unentangled states. In the ideal Wannier-Stark localization 
scenario~\cite{bhakuni:2019,emin:1987} with an infinitely strong tilted potential 
($\Delta \gg J$) and zero on-site Coulomb interaction ($U=0$), the nonergodic property can 
be represented by time-evolved quantities, including the full-chain fidelity 
$\mathcal{F}(t)=1$, the half-chain entropy $\mathcal{S}(t)=0$, and the imbalance 
$\mathcal{I}(t)=\mathcal{I}(0)$ and 
$\mathcal{I}^{\uparrow(\downarrow)}(t)=\mathcal{I}^{\uparrow(\downarrow)}(0)$ 
for arbitrarily long time. (Numerical verification for the near-ideal Wannier-Stark
localization is described in Appendix~\ref{subsec:near_ideal_WS} for $\Delta=100J$ and 
$U=0J$.) As a result, the average quantity over each episode with the maximum time horizon 
$\mathcal{T}$ should satisfy 
\begin{align} \nonumber
\langle \mathcal{F}\rangle_{\mathcal{T}} & = 1, \\ \nonumber 
\langle \mathcal{S}\rangle_\mathcal{T} &=0, \\ \nonumber
\langle \mathcal{I}\rangle_{\mathcal{T}} &= \mathcal{I}(0), \ \mbox{and} \ \\ \nonumber
\langle \mathcal{I}^{\uparrow(\downarrow)}\rangle_{\mathcal{T}} & = \mathcal{I}^{\uparrow(\downarrow)}(0).
\end{align}
These physical quantities serve as nonergodic metrics, delineating the deviation from the 
truth. Evaluating the performance of the trained DRL agent relies on its capability to 
maintain quantum nonergodicity over an extended time horizon, which involves measuring 
deviations in the nonergodic metrics.

In principle, the full-chain fidelity encapsulates the full quantum information about the 
quantum state. However, partial observation can be more efficient and feasible in 
experimental settings. For instance, the 1D tilted Bose-Hubbard model has been successfully 
realized in superconducting processors~\cite{guo:2021}. Moreover, quantum tomography 
measurements in superconducting qubits allow for direct acquisition of the elements of the 
reduced density matrix~\cite{zhang:2023}, enabling observations such as the half-chain 
entropy and sub-chain fidelity. In optical lattices, ultracold fermions can be controlled 
by a magnetic field to simulate the 1D tilted Fermi-Hubbard model with spin-resolved 
imbalance~\cite{scherg:2021}. These developments make nonergodic metrics accessible 
in experiments.

The nonergodic metrics are expected to iteratively approach the nonergodic truths during 
the training phase and demonstrate the retention of the initial state memory during the 
testing phase. In the training phase with the time horizon $\mathcal{T}=5\tau$, the 
observation results in two distinct tasks: the $\mathcal{F}_{\textnormal{sub}}$ task 
with $\mathcal{N}_{\textnormal{sub}}=1$ or $4$ sites and the 
$[\mathcal{I}^{\uparrow},\mathcal{I}^{\downarrow}]$ task. 
Figures~\ref{fig:numLattice_14_T_5_F_sub_imbalance}(a) and
\ref{fig:numLattice_14_T_5_F_sub_imbalance}(b) show, respectively the convergence of the 
sub-chain fidelity $\langle \mathcal{F}\rangle_{\mathcal{T}}$ and the spin-resolved average 
imbalance $\langle\mathcal{I}^{\uparrow(\downarrow)}\rangle_{\mathcal{T}}$.
Figure~\ref{fig:numLattice_14_T_5_F_sub_imbalance}(c) shows that a convergence of the 
consequent full-chain fidelity $\langle \mathcal{F}\rangle_{\mathcal{T}}$ has been achieved, 
agreeing with the behaviors in Figs.~\ref{fig:numLattice_14_T_5_F_sub_imbalance}(a) 
and \ref{fig:numLattice_14_T_5_F_sub_imbalance}(b). During the testing phase for 
$0<t\leq 5\tau$, the full-chain fidelity $\mathcal{F}(t)$ and half-chain entropy 
$\mathcal{S}(t)$, and the imbalances $\mathcal{I}(t)$ and 
$\mathcal{I}^{\uparrow(\downarrow)}(t)$ exhibit oscillations about their respective 
nonergodic truths, as shown in 
Figs.~\ref{fig:numLattice_14_T_5_F_sub_imbalance}(d), 
\ref{fig:numLattice_14_T_5_F_sub_imbalance}(g), and 
\ref{fig:numLattice_14_T_5_F_sub_imbalance}(h). 
The oscillatory behavior is originated from Bloch oscillations~\cite{scherg:2021,ye:2023} 
and the optimal control protocol, as shown in 
Figs.~\ref{fig:numLattice_14_T_5_F_sub_imbalance}(e) and 
~\ref{fig:numLattice_14_T_5_F_sub_imbalance}(f). The full-chain fidelity closely 
approaches the nonergodic truth value during the training, where quantum many-body 
thermalization is greatly suppressed, as can be seen from evolution of the half-chain 
entropy. The spin-resolved imbalances also oscillate about the nonzero initial value with 
a small amplitude. The three nonergodic metrics, distinguished by three different colors, 
exhibit comparable behaviors, indicating similar performance for the 
$\mathcal{F}_{\textnormal{sub}}$ and $[\mathcal{I}^{\uparrow},\mathcal{I}^{\downarrow}]$ 
tasks. In the untrained region $5\tau<t\leq 10\tau$, the nonergodic metrics oscillate more 
wildly with a slightly increased amplitude as compared to the trained time region in 
Figs.~\ref{fig:numLattice_14_T_5_F_sub_imbalance}(d), 
\ref{fig:numLattice_14_T_5_F_sub_imbalance}(g), and 
\ref{fig:numLattice_14_T_5_F_sub_imbalance}(h), implying the potential role of time 
prediction and controllability of DRL in the unexplored region.

\begin{figure*} [ht!]
\centering
\includegraphics[width=0.8\linewidth]{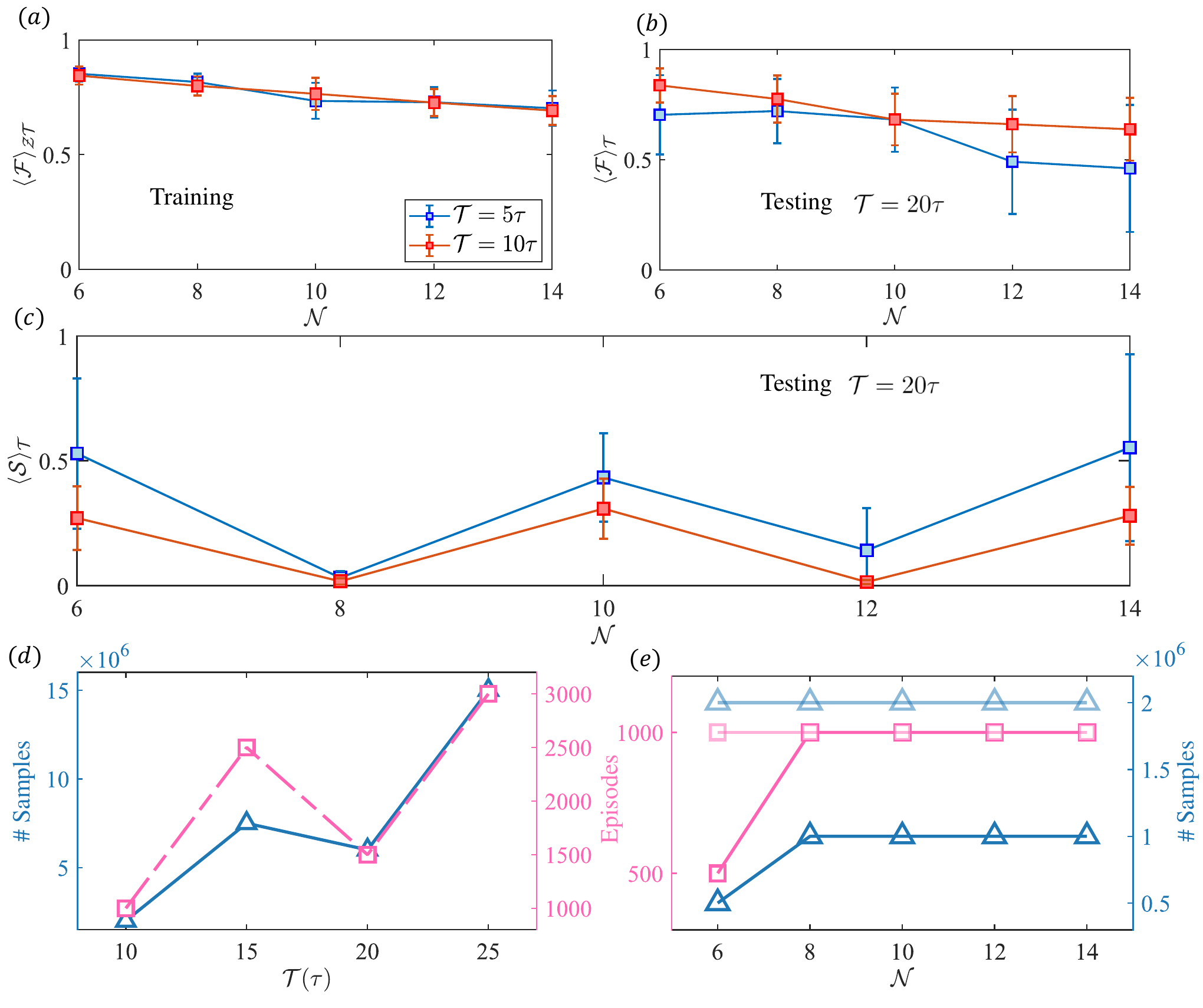}
\caption{Scalability and complexity of DRL performance and training. The observation is 
full-chain fidelity for lattice sizes $\mathcal{N} = [6, 8, 10, 12, 14]$ with the initial 
state $|-+\rangle$. (a) The training outcomes, denoted by 
$\langle \mathcal{F}\rangle_{\mathcal{ZT}}$, are assessed through $\mathcal{Z}=10$ episodes 
and calculated from the last ten episodes of the training phase. This metric provides a 
quantitative measure of the performance of the PPO agent in multiple training sessions. Two 
different training time horizons are also used: $\mathcal{T}=5\tau$ and $10\tau$. The 
average full fidelity $\langle \mathcal{F}\rangle_{\mathcal{ZT}}$ and the standard 
deviation characterize the scalability and stability of the learning process. (b,c) Testing 
results for the time horizon $\mathcal{T}=20\tau$ (including the unexplored horizon), in 
terms of the full-chain fidelity and half-chain entropy. The PPO agent is trained under two 
different time horizons: $\mathcal{T}=5\tau$ and $10\tau$. The error bars denote the 
standard deviation within a single episode, providing a measure of the variability and 
reliability of the agent's performance across different training duration. (d) Worst-case 
sample complexity and the corresponding training episodes across various training time 
horizons. The worst-case scenario are represented by the blue curves and corresponding axes, 
which is from seven distinct initial states that include $|-+-+\rangle$, $|+-+-\rangle$, 
and five other permutations of spin configurations: 
$|\uparrow\downarrow\uparrow\downarrow\uparrow\downarrow\uparrow\downarrow\rangle$, 
$|\uparrow\downarrow\uparrow\uparrow\downarrow\downarrow\uparrow\downarrow\rangle$, 
$|\downarrow\uparrow\downarrow\uparrow\downarrow\uparrow\downarrow\uparrow\rangle$, 
$|\uparrow\downarrow\uparrow\downarrow\uparrow\uparrow\downarrow\downarrow\rangle$, 
and $|\uparrow\uparrow\downarrow\downarrow\uparrow\uparrow\downarrow\downarrow\rangle$. 
The training episodes are determined by assessing the qualitative convergence of the 
full-chain fidelity $\langle \mathcal{F}\rangle_{\mathcal{T}}$ for a chain of size 
$\mathcal{N}=8$, as represented by the pink curve and axis. The neural network 
configuration is {\bf Config A}. (e) Sample complexity and the corresponding training 
episodes versus lattice chain size for the fixed initial state $|-+\rangle$. The 
full-chain fidelity varies distinctly across two time horizons. For $\mathcal{T}=5\tau$, 
the DRL employs the neural network configuration of {\bf Config B}, shown by dark blue 
and pink curves. For a longer time horizon of $\mathcal{T}=10\tau$, the training DRL 
utilizes the {\bf Config A}, as illustrated by light blue and pink curves.} 
\label{fig:scalability_sample_complexity}
\end{figure*}

\subsection{Scalability and complexity of deep reinforcement learning}

In general, RL deals with sequential decision-making problems, so the complexity 
of the PPO agent algorithm involves not only the number of samples but also the 
quality and variety of the quantum environment that the agent encounters. For the 
tilted Fermi-Hubbard chain, the lattice size determines the dimension of the Hilbert 
space in which rich quantum states or phases arise. A longer time horizon increases
the time complexity for the PPO agent. When applied to a quantum many-body system, 
a key attribute of the PPO agent algorithm is the scalability of performance with 
the lattice size, training time complexity and the sample number. These factors 
can directly influence the feasibility of the DRL control in larger systems. To
study the scalability, we define the number of samples as the sample complexity,
in which the time complexity can be directly encoded, and systematically test the 
performance for various lattice sizes and training time horizons. The results are
shown in Fig.~\ref{fig:scalability_sample_complexity}. Specifically, 
Figs.~\ref{fig:scalability_sample_complexity}~(a-c) present the results for 
lattice sizes $\mathcal{N}=[6,8,10,12,14]$, demonstrating stable performance of 
DRL even in the unexplored time horizons and suggesting the feasibility of extending 
DRL nonergodicity control into a larger Hilbert space. For the worst-case sample 
complexity, i.e., the number of observation points at each time over the whole 
training process, is about $10^6-10^7$ for the seven tested initial states in 
Fig.~\ref{fig:scalability_sample_complexity}(d). Especially, the sample complexity 
shows that the performance is largely independent of the lattice size, as a result
of the nonergodicity control mechanism, i.e., the tendency to gradually converge 
to the approximated control in the single-particle picture.

\section{Physical interpretation and robustness of DRL} \label{sec:physics}

To understand the mechanism of quantum nonergodicity control, we recall the 
phenomenon of Anderson localization~\cite{anderson:1958,vsuntajs:2023,thompson:2010}
in the single-particle picture. It arises from independent and identical random 
chemical potentials assigned to each lattice site described by the tight-binding 
model. The disorders characterized by the magnitude of the range in the random 
on-site potential, if sufficiently strong, will disrupt the quantum ergodicity. 
Single-particle localization can also occur without disorders. For example, 
substituting the random potential with a uniform electric field can lead to 
Wannier-Stark localization~\cite{bhakuni:2019,emin:1987}. This effect has been
observed under a sufficiently strong tilted potential in a superconducting 
quantum processor~\cite{guo:2021}. 

\begin{figure*} [ht!]
\centering
\includegraphics[width=0.8\linewidth]{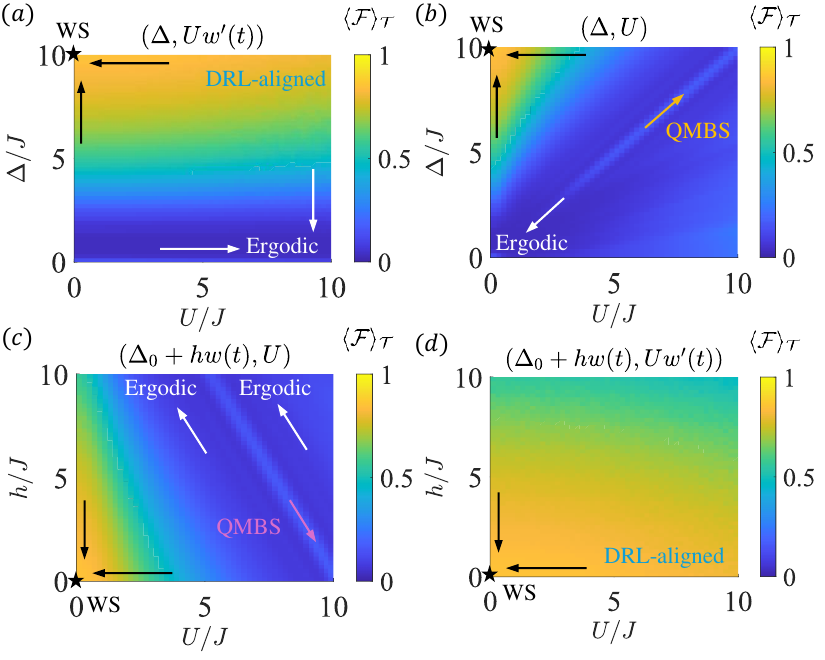}
\caption{Physical interpretation and robustness of DRL for quantum 
nonergodicity control in terms of the phase diagrams of the fidelity.
(a-d) Control parameter space for the full-chain fidelity 
$\langle \mathcal{F}\rangle_{\mathcal{T}}$ in different parameter planes: 
$(\Delta,Uw^{\prime}(t))$, $(\Delta, U)$, $(\Delta_0+hw(t),U)$, and 
$(\Delta_0+hw(t),Uw^{\prime}(t))$, where the random numbers $w(t)\in[-1,0]$ and 
$w^{\prime}(t)\in[-1,1]$ are independent, identical, and uniformly distributed 
at each time step for a fixed $\Delta_0=10J$, and constants $\Delta$, $U$, and 
$h$. The fidelity $\langle \mathcal{F}\rangle_{\mathcal{T}}$ is calculated using 
$\mathcal{Z}=100$ disorder averages, with the system initialized in the state 
$|-+-+\rangle$ in a lattice of size $\mathcal{N}=8$ over the time horizon 
$\mathcal{T}=10\tau$ that matches the testing time horizon of DRL in 
Fig.~\ref{fig:numLattice_14_T_5_F_sub_imbalance}.} 
\label{fig:phase_diagram}
\end{figure*}

The picture of single-particle localization provides insights into the phenomenon
of MBL~\cite{abanin:2019} - the augmentation of Anderson localization 
with constant on-site Coulomb interaction~\cite{sierant:2018}. It was also found
that quasirandom disorders in the chemical potentials in the Aubry-Andre 
model~\cite{schreiber:2015,vu:2022,kohlert:2019} can lead to MBL. Moreover, 
a strong random Coulomb interaction at each site has been demonstrated to 
facilitate the onset of MBL~\cite{lev:2016,sierant:2017,sierant:2018}. The spatial 
distribution of the electric field further facilitates MBL in the presence of 
many-body interactions. For instance, Stark many-body
localization~\cite{schulzP:2019} emerges under a nonuniform electric field when 
a harmonic term is present that breaks the pure linearity of the electric field. 
In addition, sufficiently strong random fields superimposed on a uniform electric 
field can trigger MBL~\cite{van:2019}. It is worth noting that MBL resides within 
the realm of strong ergodicity breaking, whereas QMBS corresponds to weak 
ergodicity breaking, both violating the eigenstate thermalization 
hypothesis~\cite{deutsch:1991,srednicki:1994,rigol:2008}, where the quantum dynamics 
of QMBS depend on the initial conditions rooted in the disconnected structure in 
the Hilbert space.

Deep RL delivers optimal control to induce nonergodicity in a quantum many-body 
system. Figures~\ref{fig:numLattice_14_T_5_F_sub_imbalance}(e) and 
\ref{fig:numLattice_14_T_5_F_sub_imbalance}(f) show the optimally controlled 
trajectories for $\Delta(t)$ and $U(t)$, respectively. For the control based
on $\Delta(t)$, it tends to converge to a constant value: either $\Delta/J=10$ or 
$-10$, where the sign is due to the different titled directions within the 
lattice chain. Alternatively, incorporating random perturbations into the constant 
$\Delta$ also represents a potential control protocol by DRL. For optimal 
protocol based on $U(t)$, it oscillates within the original search range: 
$-10\leq U(t)/J\leq 10 $. To understand the quantum phases that DRL learns 
and why it converges to some specific values as exemplified in 
Figs.~\ref{fig:numLattice_14_T_5_F_sub_imbalance}(e) and 
\ref{fig:numLattice_14_T_5_F_sub_imbalance}(f), 
we simplify the optimal protocol of $\Delta(t)$ as $\Delta(t)/J=\Delta_0/J=10$,
where $\Delta_0$ is a constant, or incorporates random perturbation as 
$\Delta(t)=\Delta_0+hw(t)$ with the constant $h/J\in[0,10]$ and random number 
$w(t)\in[-1,0]$ at each time step, ensuring it stays within the range 
$\Delta(t)/J\in[0,10]$ as in Fig.~\ref{fig:numLattice_14_T_5_F_sub_imbalance}(e).
For the optimal protocol of $U(t)$, we simplify it as $U(t)=Uw^{\prime}(t)$ with 
the constant $U/J\in[0,10]$ and random numbers $w^{\prime}(t)\in[-1,1]$ at each time
step, limiting $U(t)/J$ between $-10$ and $10$. For convenience, we use the term
`Deep RL-aligned protocol'' to denote the simplified control protocols. The quantum 
phase generated by the DRL-aligned protocol is referred to as the 
``Deep RL-aligned phase''. Comparing the DRL-aligned and other quantum phases, 
especially the QMBS with the constant $\Delta\approx U\gg J$, entails testing  
four pairs of actions: $(\Delta, Uw^{\prime}(t))$, $(\Delta, U)$, 
$(\Delta_0+hw(t),U)$, and $(\Delta_0+hw(t),Uw^{\prime}(t))$, as illustrated in 
Fig.~\ref{fig:phase_diagram}.

We use the average fidelity $\langle \mathcal{F}\rangle_{\mathcal{T}}$ for the 
whole chain to interpret DRL and characterize the quantum phases, which is
justified, as follows. In episodic learning, the DRL agent collects a sequence 
of observations, rewards, and actions within each episode, subsequently updating 
its policy to maximize the accumulated reward for future training. Observations 
could consist of fidelity or imbalance, with the reward function directly linked to 
the observation and aimed at converging to its initial value. As a result, DRL 
is designed to focus solely on minimizing the discrepancy between the accumulated 
fidelity or imbalance and its initial value. Plotting the average fidelity over 
one episode offers a way to understand the physical mechanisms of learning with 
DRL. While imbalance only provides partial information about the quantum state, 
fidelity encompasses the entire chain, making it an appropriate indicator. We note 
that the average fidelity was used to characterize the quantum phases in a previous 
work~\cite{van:2019}. 

\begin{figure} [ht!]
\centering
\includegraphics[width=\linewidth]{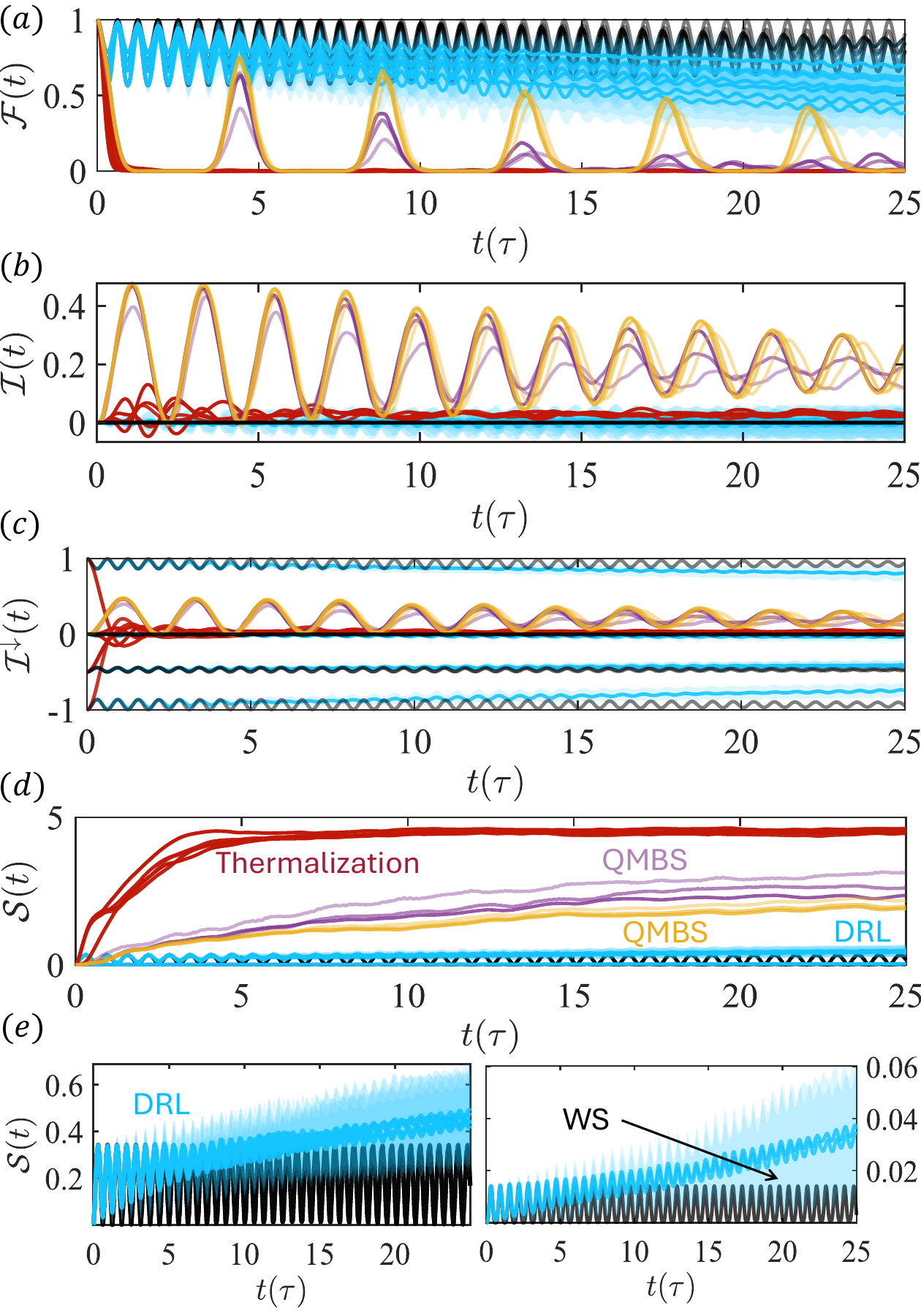}
\caption{Comparison of DRL control with other protocols in terms of the quantum 
phases by using time-dependent nonergodic metrics. These metrics include the 
(a) full-chain fidelity $\mathcal{F}(t)$, (b) imbalance $\mathcal{I}(t)$, 
(c) spin-down imbalance $\mathcal{I}^{\downarrow}(t)$, and (d,e) the half-chain 
Von Neuman entropy $\mathcal{S}(t)$. The analysis is conducted over the time 
horizon of $\mathcal{T}=25\tau$ in a lattice of size $\mathcal{N}=8$. Beginning 
with the QMBS state $|-+\rangle$~\cite{desaules:2021}, the QMBS phase is 
illustrated by the yellow curves for $\Delta/J=U/J=6,8,10$ (from lighter to 
darker shades). A random perturbation yields the QMBS state in purple for
$(\Delta_0+hw(t),U) = (10+10w(t),5.2)J,(10+5.8w(t),7.3)J, (10+0.5w(t),10)J$,
displayed in progressively darker curves. Also shown are DRL results and 
two other quantum phases, the Wannier-Stark and thermalization phases.
Seven initial states are used, including the typical QMBS, 
$|-+-+\rangle$, $|+-+-\rangle$ and five other permutations of spin configurations: 
$|\uparrow\downarrow\uparrow\downarrow\uparrow\downarrow\uparrow\downarrow\rangle$, 
$|\uparrow\downarrow\uparrow\uparrow\downarrow\downarrow\uparrow\downarrow\rangle$, 
$|\downarrow\uparrow\downarrow\uparrow\downarrow\uparrow\downarrow\uparrow\rangle$, 
$|\uparrow\downarrow\uparrow\downarrow\uparrow\uparrow\downarrow\downarrow\rangle$, 
and $|\uparrow\uparrow\downarrow\downarrow\uparrow\uparrow\downarrow\downarrow\rangle$ (plotted in identical colors). 
The Wannier-Stark phase is characterized by constants $\Delta/J = 10$ and $U/J=0$, 
while thermalization features constants $\Delta/J=1$, $U/J=2$. The three quantum 
phases serve as benchmarks for evaluating the performance of DRL. The blue 
curves depict the testing outcomes of DRL, averaged over $\mathcal{Z}=100$ 
independent testing episodes. The standard deviations are also shown. Deep RL is 
trained to learn the quantum system using the full-chain fidelity observable with 
the training time of $\mathcal{T}=25\tau$ and neural network size of {\bf Config A} as 
described in Sec.~\ref{subsec:PPO}.} 
\label{fig:fullFidelity_T_25_RL_compar_benchmark}
\end{figure}

To understand why the optimal protocol converges to a specific region, as 
demonstrated in Figs.~\ref{fig:numLattice_14_T_5_F_sub_imbalance}(e) and 
\ref{fig:numLattice_14_T_5_F_sub_imbalance}(f), we examine the quantum phase 
diagram to reveal what the DRL agent has learned for nonergodicity control 
tailored to the quantum many-body system. The testing time horizon $\mathcal{T}$ 
dictates the temporal span for constructing the quantum phase diagram of 
$\langle \mathcal{F}\rangle_{\mathcal{T}}$. Despite the short observation time, 
the phase diagram highlights the DRL-aligned regime and distinct quantum 
phases, as illustrated in Fig.~\ref{fig:phase_diagram} for the full lattice model, 
where QMBS, thermalization or ergodic, and Wannier-Stark phases are displayed. More 
details are shown by the related time-dependent nonergodic metrics in 
Fig.~\ref{fig:fullFidelity_T_25_RL_compar_benchmark}.

For $\Delta\approx U\gg J$, the first-order Schrieffer-Wolff 
transformation~\cite{bravyi:2011} is applicable, which can be used to derive   
the effective model of the full lattice system~\cite{desaules:2021}. The QMBS states 
$|-+\rangle$ and $|+-\rangle$ associated with weak ergodic breaking exhibit switching 
dynamics within a hypergrid structure characterized by the time-dependent nonergodic 
metrics, as described in a previous work~\cite{desaules:2021}. The inherent tower 
structure of the overlap between QMBS states and the eigenstates~\cite{desaules:2021}
is indicative of a concentration about some specific energy levels and violation 
of eigenstate thermalization hypothesis. In our work, we use the full lattice model 
with periodic boundary conditions and the approximation $\Delta\approx U \approx 10 J$ 
to find QMBS states with behavior similar to that of the corresponding states in the 
effective model, as shown in Figs.~\ref{fig:phase_diagram}(b,c) and 
Figs.~\ref{fig:fullFidelity_T_25_RL_compar_benchmark}(a-d). The phase diagram of 
$\langle\mathcal{F}\rangle_{\mathcal{T}}$ in Fig.~\ref{fig:phase_diagram}(b) 
reveals the presence of the QMBS phase along the direction indicated by the orange 
arrow, with a darker blue area nearby. This finding agrees with an earlier 
result~\cite{desaules:2021} on the $U-\Delta$ phase diagram featuring the first 
peak of the imbalance. Fig.~\ref{fig:phase_diagram}(c) reveals that random 
perturbations to the tilted potential result in a similar QMBS pattern, suggesting 
the robustness of the QMBS states. Additional results supporting the robustness
are shown by the time-dependent nonergodic metrics in 
Figs.~\ref{fig:fullFidelity_T_25_RL_compar_benchmark}(a-d), where the 
perturbations lead to a quicker decay of the revival amplitude in the full-chain 
fidelity and imbalance, though with a slight rise in the entanglement entropy.

Within the time horizon $\mathcal{T}=10\tau$, there is a tendency towards ergodic 
behavior in the quantum dynamics, as indicated by the white arrows in the phase 
diagrams in Figs.~\ref{fig:phase_diagram}(a-c). For $\Delta/J=1$ and $U/J=2$, 
the short-time thermalization process is demonstrated in 
Figs.~\ref{fig:fullFidelity_T_25_RL_compar_benchmark}(a-d). There is a swift 
decline and stabilization in the fidelity and imbalance, accompanied by a rapid 
convergence of the entanglement entropy. The Wannier-Stark phase can usually be 
characterized by $\Delta\gg J, U$ but, due to the limited action range, the action 
pair of $\Delta=10J$ and $U=0J$ emerges as the closest approximation to the ideal 
Wannier-Stark phase, as shown in both Figs.~\ref{fig:phase_diagram} and 
\ref{fig:fullFidelity_T_25_RL_compar_benchmark}. The regions in the vicinity 
of the Wannier-Stark point in the phase diagrams exhibit a tendency towards the 
Wannier-Stark phase, as indicated by the black arrows in 
Figs.~\ref{fig:phase_diagram}(a-d). In terms of the time-dependent nonergodic 
metrics, the Wannier-Stark phase serves as an ideal benchmark for initial memory 
retention and nonergodicity control. It maintains the initial values of the 
fidelity, imbalance, and entropy, in spite of the Bloch oscillations~\cite{ye:2023} 
of the period $t_{B} = 2\pi/\Delta$. Within the search space
$\Delta/J\in[-10,10]$ and $U/J\in[-10,10]$, the Wannier-Stark phase can simply
be regarded as a specific point at $\Delta/J=10$ and $U/J=0$, which is sensitive 
to constant perturbations from the on-site Coulomb interaction $U$ as indicated
in Figs.~\ref{fig:phase_diagram}(b) and \ref{fig:phase_diagram}(c) but is 
robust against perturbations $hw(t)$ in the tilted potential $\Delta_0$ to some 
extent, as shown in Figs.~\ref{fig:phase_diagram}(c) and \ref{fig:phase_diagram}(d).
The DRL-aligned protocol also reveals the robustness against perturbations 
for $Uw^{\prime}(t)$, as shown in Figs.~\ref{fig:phase_diagram}(a) and 
\ref{fig:phase_diagram}(d), highlighting a broad control scheme where DRL 
converges to maximize the accumulated fidelity.

While the results in Fig.~\ref{fig:phase_diagram} are from the DRL-aligned protocol, 
the real DRL control flow and the corresponding performance of the nonergodic 
metrics are demonstrated in Fig.~\ref{fig:fullFidelity_T_25_RL_compar_benchmark}.
In particular, Fig.~\ref{fig:fullFidelity_T_25_RL_compar_benchmark}(e) reveals a
consistency between the DRL protocol and Wannier-Stark fidelity in the short 
term, exhibiting the same period of Bloch oscillations. However, in the long run,
a slight deviation in the nonergodic metrics emerges, representing the trade-off 
between robustness and performance of nonergodicity control in the DRL
action protocol.

\begin{figure*} [ht!]
\centering
\includegraphics[width=\linewidth]{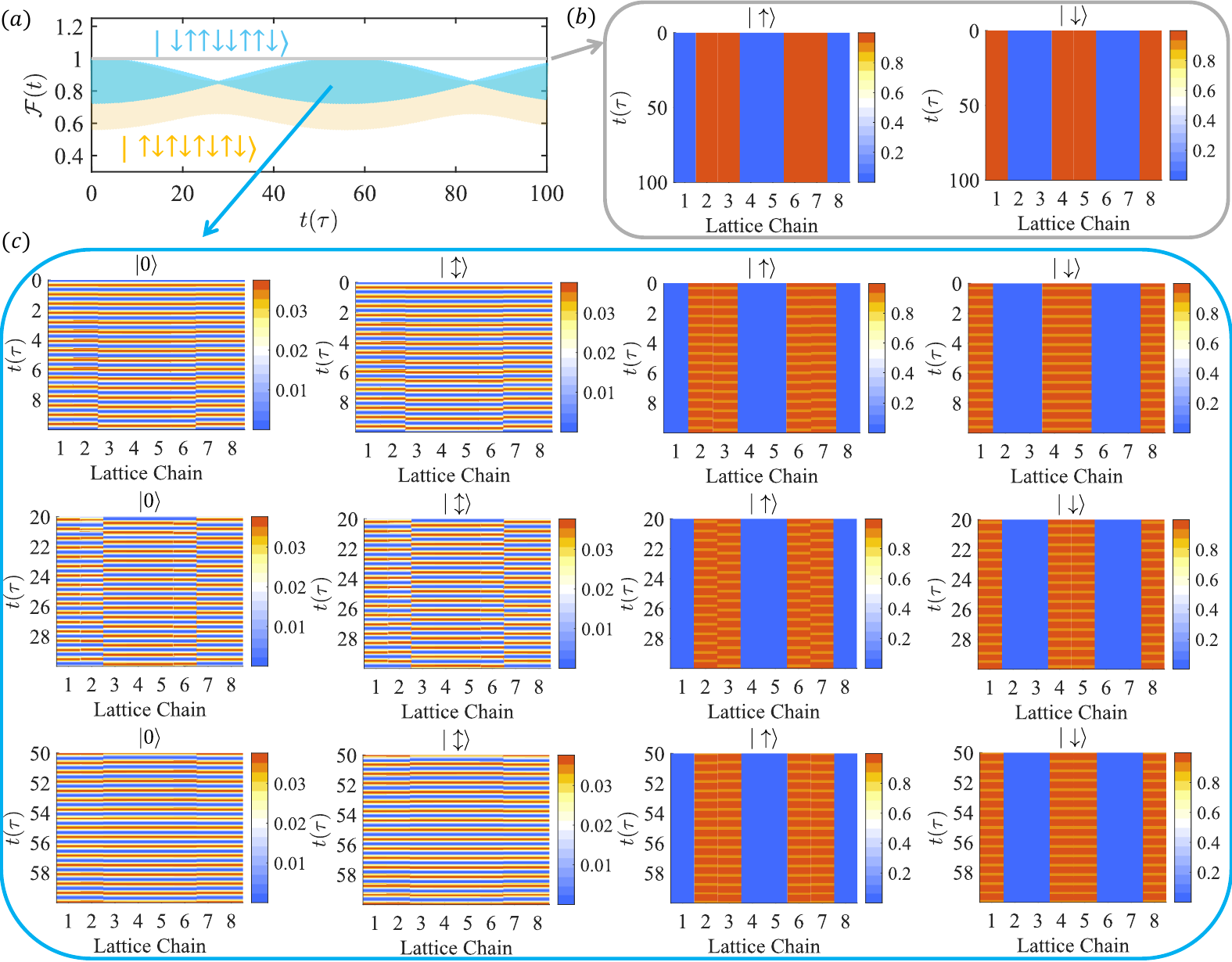}
\caption{Interpretation of the envelope oscillation of the Wannier-Stark fidelity 
in a lattice chain with $\mathcal{N}=8$ sites. (a) Periodic oscillations in the 
envelope amplitude of the time-dependent fidelity of the full lattice chain,
along with the short-time Bloch oscillation. The grey and blue curves represent 
simulations starting from the same initial state $|-+-+\rangle$, with different 
action pairs: $(\Delta,U) = (100, 0)J$ and $(10,0)J$, respectively. The yellow 
curves are initialized with another state 
$|\uparrow\downarrow\uparrow\downarrow\uparrow\downarrow\uparrow\downarrow\rangle$, 
with the action pair $(\Delta,U) = (10,0)J$. (b) Ideal Wannier-Stark phase 
visualized by the probability distribution in the 2D parameter plane spanning 
space and time. (c) Time slice of the space-time panel for
$t\in[0,10]\tau,[20,30]\tau,[50,60]\tau$, which displays the probability 
distribution in the basis $|0\rangle$, $|\updownarrow\rangle$, $|\uparrow\rangle$, 
and $|\downarrow\rangle$ at each site location.} 
\label{fig:WS_initial_minus_plus}
\end{figure*} 

The control method can be extended to training over longer time horizons, where
some unexplored characteristics of the Wannier-Stark phase can be revealed. Over 
an extended period, the Wannier-Stark phase exhibits both short-term Bloch 
oscillations, as illustrated in Fig.~\ref{fig:fullFidelity_T_25_RL_compar_benchmark},
and envelope oscillations in the time-dependent series, as shown in 
Fig.~\ref{fig:WS_initial_minus_plus}(a). The period of the envelope oscillations 
is positively correlated with the tilted potential, while the period of the Bloch 
oscillations has a negative correlation with it. The amplitude of the Bloch 
oscillations and the shape of the envelope oscillations are influenced by the 
initial state conditions, as illustrated in 
Figs.~\ref{fig:fullFidelity_T_25_RL_compar_benchmark}(e) and 
\ref{fig:WS_initial_minus_plus}(a), respectively.
The ideal Wannier-Stark phase is shown in Fig.~\ref{fig:WS_initial_minus_plus}(b),
which can be used as a benchmark. Figure~\ref{fig:WS_initial_minus_plus}(c) reveals 
that the spatial oscillations of a quantum state contribute to the envelope 
oscillation. In general, simulating quantum many-body systems is challenging 
due to the difficulty of exponential growth in the computational complexity with
the system size, but experiments are possible. With the numerical validation of 
the feasibility of DRL for controlling a small system, the training methodology 
can in principle be extended to experiments with larger systems.

\section{Discussion} \label{sec:discussion}

In complex quantum many-body systems, thermalization leading to ergodicity is 
a major source of decoherence. Developing methods to suppress thermalization to
achieve nonergodicity is essential for applications, e.g., in quantum information
science and technology. AI-based optimal control provides the possibility of 
controlling complex quantum many-body systems to achieve nonergodicity. Utilizing 
the 1D tilted Fermi-Hubbard model as a paradigm, we developed a model-free DRL 
approach to controlling quantum nonergodicity, where the DRL agent interacts with 
the quantum environment stipulated by the quantum many-body system. During the 
online training phase, the DRL agent, specifically a PPO agent, collects time 
series data in real time within one episode, which include observations, rewards, 
and actions. A criterion to choose the type of observations is their experimental 
accessibility in real time. For the 1D tilted Fermi-Hubbard chain, both the full or 
partial chain fidelity~\cite{guo:2021,zhang:2023} and the spin-resolved 
imbalance~\cite{scherg:2021} satisfy this criterion. The reward values are 
determined by the observable variable and a function tailored to meeting the 
nonergodic objective that the metrics preserve the initial memory over a long 
time. Consequently, the reward function is crafted to maintain the time-dependent 
nonergodic metric as close to its initial value as possible. The stochastic policy 
dictates subsequent actions, $\Delta(t)$ and $U(t)$ for the tilted Fermi-Hubbard chain, 
can physically be implemented through a properly designed magnetic 
field~\cite{scherg:2021,kohlert:2023,schreiber:2015} within an episode. This 
policy is a stochastic probability distribution over all possible actions 
conditioned on the given observable at that time. The policy is updated after 
each episode to maximize the accumulated reward over one episode. The well-trained 
PPO agent is saved after a predetermined number of training episodes designed to 
ensure the complete convergence of the mean reward training curve. In the online 
testing phase during which the optimal policy determined by training is not updated, 
the well-trained PPO agent applies the optimal nonergodic control to the same 
quantum system in real time. The optimal policy of the well-trained PPO agent 
gives the subsequent actions based on its observations. The control is completely
data-driven in the sense that, in the whole training and testing process, no prior 
knowledge about the target quantum many-body system is required: all needed is 
model-free DRL with experimentally available observables. The quantum phases that 
the DRL agent has learned can be used to understand the physical mechanisms underlying 
the optimal control policy. 

Two patterns arising from the DRL control are that, initially, the absolute 
value of the titled potential term tends to reach a constant maximum value, while 
the on-site Coulomb interaction appears random within the original search range 
(DRL-aligned policy). We explored all possible combinations of these two patterns and 
discovered a rich array of quantum phases through various phase diagrams, including 
ergodic, QMBS, and Wannier-Stark phases. Both the QMBS and Wannier-Stark phases are 
robust against perturbations to the tilted potential, but the latter is sensitive to 
the constant on-site Coulomb potential and tends to thermalize. In contrast, the 
DRL-aligned policy offers a broad control scenario for perturbations to either the 
tilted term or the on-site Coulomb interaction. By comparing these phases with the 
actual DRL policy over time series, we observed that the DRL policy closely aligns 
with the Wannier-Stark phase. A physical analysis indicates that, under the condition 
of an infinite tilted term, the Wannier-Stark phase approaches the ideal state, which 
is indicative of single-particle localization. This provides a simplistic protocol for 
our control task. In general, the DRL protocol offers superior control robustness, with 
performance comparable to the Wannier-Stark phase under the nonergodicity control 
objective. Another appealing feature of DRL nonergodicity control is that the 
observations from even just one site suffice for realizing the control goal, 
facilitating experimental implementation. 

Recent work~\cite{lu:2024} employs a model-based machine learning approach—specifically, the variational entanglement-enhancing field—to optimize the magnetic field in quantum spin chains, enabling persistent ballistic entanglement spreading. By adjusting local parameters in the Hamiltonian, this approach accelerates the saturation of ergodicity beyond what would typically occur in a homogeneous, time-independent Hamiltonian. In a certain sense, these results stand in contrast to our work: while they seek to expedite ergodicity from an initial product state, we aim to prevent its onset. This highlights the potential of time-dependent control fields to enable a wide range of physical implementations, whether for accelerating or suppressing ergodicity. The application of machine learning to optimize control fields in quantum systems provides a powerful and versatile tool for tackling more complex quantum control tasks.
\section*{Data and code availability}

The data and code for this work are available at:
GitHub: https://github.com/liliyequantum/nonergodicityRL.

\section*{Acknowledgments}

We thank Prof.~L. Ying for comments and suggestions, Dr.~J.-L. Wang for a discussion
on the entanglement-entropy calculation, and Dr.~A. Yang for general discussions on 
quantum many-body systems. We also thank the group of Prof.~M. Aidelsburger for 
providing their published codes. This work was supported by the Air Force Office of 
Scientific Research under Grant No.~FA9550-21-1-0438 and by the Office of Naval Research 
under Grant No.~N00014-24-1-2548.

\appendix

\section{1D tilted Fermi-Hubbard model} \label{appendix_A}

\subsection{Fock basis}

The behavior of interacting fermions, constrained to move along a 1D lattice and 
interacting via the on-site Coulomb interaction, is described by the 1D tilted 
Fermi-Hubbard model in the presence of an external linear potential. Here, fermions 
are represented as spin $1/2$ particles with spin up (down), with the respective
numbers $\mathcal{N}_{\uparrow}$ and $\mathcal{N}_{\downarrow}$ among the 
$\mathcal{N}$ lattice sites. Accordingly, the number of bases for the spin up (down), 
denoted by $d_{\uparrow}$ ($d_{\downarrow}$), is determined as
\begin{equation}\label{eq:basis}
d_{\uparrow(\downarrow)}=\left[\begin{array}{c} \mathcal{N}\\\mathcal{N}_{\uparrow(\downarrow)}\end{array}\right].
\end{equation}
In the Fock space, the entire basis is constructed by combining the basis states 
for spin up and spin down:
$$\hat{c}^{\dagger}_{i_1}\hat{c}^{\dagger}_{i_2}...\hat{c}^{\dagger}_{i_{\mathcal{N}_{\uparrow}}}\hat{c}^{\dagger}_{j_1}\hat{c}^{\dagger}_{j_2}...\hat{c}^{\dagger}_{j_{\mathcal{N}_{\downarrow}}}|0\rangle,$$
which corresponds to one-to-one pairs of tuples:
\begin{equation}
    ((i_1,i_2,...,i_{\mathcal{N}_{\uparrow}}),(j_{1},j_{2},...,j_{\mathcal{N}_{\downarrow}}))=(\alpha,\beta),
\end{equation}
where a symbol, such as $i_{\mathcal{N}_{\uparrow}}$ or $j_{\mathcal{N}_{\downarrow}}$,
records the occupied site location of the $\mathcal{N}_{\uparrow}$-th spin up or 
$\mathcal{N}_{\downarrow}$-th spin down particle at the lattice site.

The total number of possible pairs of tuples is $d_{\uparrow}\times d_{\downarrow}$, 
representing the Hilbert space dimension in the 1D tilted Fermi-Hubbard model with 
particle-number conservation. The general quantum state over the 1D spin-lattice 
can be expanded in the Fock basis as
\begin{equation}
    |\psi\rangle = \sum\limits_{\alpha,\beta}|\alpha,\beta\rangle\langle\alpha,\beta|\psi\rangle\equiv \sum\limits_{\alpha,\beta}M^{(\psi)}_{\alpha\beta}|\alpha,\beta\rangle.
\end{equation}
Computationally, the quantum state can be denoted by the matrix $M^{(\psi)}$ of
dimension $d_\uparrow \times d_\downarrow$.

The Hamiltonian governing the time evolution of the quantum state is
\begin{align}\nonumber
\hat{H} &= -J\sum\limits_{j}\left(\hat{c}^{\dagger}_{j+1,\uparrow}\hat{c}_{j,\uparrow}+\textnormal{h.c.}\right)\\ \nonumber
&-J\sum\limits_{j}\left(\hat{c}^{\dagger}_{j+1,\downarrow}\hat{c}_{j,\downarrow}+\textnormal{h.c.}\right)\\ \nonumber
&+U\sum\limits_{j}\hat{n}_{j,\uparrow}\hat{n}_{j,\downarrow} 
+\Delta \sum\limits_{j}j\hat{n}_{j,\uparrow}+\Delta \sum\limits_{j}j\hat{n}_{j,\downarrow}, 
\end{align}
where $\hat{c}_j$ and $\hat{c}^{\dagger}_j$ are the fermionic annihilation and 
creation operators, respectively, and $\hat{n}_j$ is the particle number operator 
distinguished by spin up and down. The parameters $J$, $U$, and $\Delta$ denote the 
nearest-neighbor hopping strength, the on-site Coulomb interaction, and the strength 
of the tilted potential, respectively. We use periodic boundary conditions so as
to maintain the continuity of the state across the boundary. For example, the 
transition of a spin-up fermion from the last site to the first site is represented 
as $\hat{c}^{\dagger}_{1,\uparrow}\hat{c}_{L,\uparrow}$. The periodic boundary 
conditions ensure that the lattice behaves as if it were looped, allowing for 
seamless transitions of particles across the boundary.

To obtain the time evolution of the Schr\"{o}dinger system, we refer to previous 
works~\cite{scherg:2021} and modify the Hamiltonian as
\begin{align}\nonumber
\hat{H} &= 
(\hat{H}^{\textnormal{hop}}_{\uparrow})_{d_{\uparrow}\times d_{\uparrow}}\otimes \mathbb{I}_{d_{\downarrow}\times d_{\downarrow}}
+\mathbb{I}_{d_{\uparrow}\times d_{\uparrow}}\otimes (\hat{H}^{\textnormal{hop}}_{\downarrow})_{d_{\downarrow}\times d_{\downarrow}}\\ \nonumber
    &+(\hat{H}^{\textnormal{diag}})_{(d_{\uparrow}\times d_{\downarrow})\times(d_{\uparrow}\times d_{\downarrow})},
\end{align}
where each spin type only hops within its respective subspace 
$\hat{H}^{\textnormal{hop}}$ and the diagonal components of 
$\hat{H}^{\textnormal{diag}}$ are represented as $V$:
\begin{equation}\nonumber
    V_{\alpha,\beta} = \sum\limits_{k=1}^{N_{\uparrow}} V_{i_{k},\uparrow}+\sum\limits_{k=1}^{N_{\downarrow}}V_{j_{k},\downarrow}+U|(i_1,i_2,...,i_{N_{\uparrow}})\cap (j_1,j_2,...,j_{N_{\downarrow}})|,
\end{equation}
where $|(i_1,i_2,...,i_{N_{\uparrow}})\cap (j_1,j_2,...,j_{N_{\downarrow}})|$ denotes 
the count of the identical elements between the two sets. Consequently, the 
Schr\"{o}dinger equation in the 1D tilted Fermi-Hubbard model is simplified to
\begin{equation}\label{eq:quantum_dynamics}
    i\hbar|\dot{\psi}\rangle = (\hat{H}^{\textnormal{hop}}_{\uparrow}\otimes\mathbb{I}_{\downarrow})|\psi\rangle + (\mathbb{I}_{\uparrow}\otimes\hat{H}^{\textnormal{hop}}_{\downarrow})|\psi\rangle + \hat{H}^{\textnormal{diag}}|\psi\rangle,
\end{equation}
where the dot over the quantum state $|\psi\rangle$ denotes its time derivative, 
specifically:
$$|\dot{\psi}\rangle\equiv\partial{|\psi\rangle}/\partial{t}=\sum\limits_{\alpha,\beta}\dot{M}^{(\psi)}_{\alpha\beta}|\alpha\beta\rangle.$$ 
Normalizing both sides of Eq.~\eqref{eq:quantum_dynamics} with the hopping strength 
$J$, we obtain a dimensionless equation. The time unit is defined as 
$\tau \equiv \hbar/J$, and the potential terms $\Delta$ and $U$ are expressed in 
units of $J$. 

The Schr\"{o}dinger equation can be recast in the matrix form, expanded in the Fock 
basis $|\alpha\beta\rangle$, through the following transformation:
\begin{align}\nonumber
\left(\hat{H}^{\textnormal{hop}}_{\uparrow}\otimes\mathbb{I}_{\downarrow}\right)|\psi\rangle &=\sum\limits_{\gamma\beta}M^{(\psi)}_{\gamma\beta}\left(\sum\limits_{\alpha}H^{\textnormal{hop}}_{\uparrow,\alpha\gamma}|\alpha\rangle_{\uparrow}\right)\otimes|\beta\rangle_{\downarrow}\\\nonumber
&=\sum\limits_{\alpha\beta}\left(H^{\textnormal{hop}}_{\uparrow}M^{(\psi)}\right)_{\alpha\beta}|\alpha\beta\rangle,
\end{align}
\begin{align}\nonumber
    \left(\mathbb{I}_{\uparrow}\otimes \hat{H}^{\textnormal{hop}}_{\downarrow}\right)|\psi\rangle&=\sum\limits_{\alpha\gamma}M^{(\psi)}_{\alpha\gamma}|\alpha\rangle_{\uparrow}\otimes\left(\sum\limits_{\beta}H^{\textnormal{hop}}_{\downarrow,\beta\gamma}|\beta\rangle_{\downarrow}\right)\\\nonumber    &=\sum\limits_{\alpha\beta}\left(M^{(\psi)}H^{\textnormal{hop}}_{\downarrow}\right)_{\alpha\beta}|\alpha\beta\rangle,
\end{align}
and
\begin{align}\nonumber
    \hat{H}^{\textnormal{diag}}|\psi\rangle&=\sum\limits_{\alpha,\beta}M^{(\psi)}_{\alpha\beta}\hat{H}^{\textnormal{diag}}|\alpha\beta\rangle\\\nonumber
    &=\sum\limits_{\alpha,\beta}\left(V\circ M^{(\psi)}\right)_{\alpha\beta}|\alpha\beta\rangle.
\end{align}
The matrix form of the Schr\"{o}dinger equation is given by:
\begin{equation}\label{eq:matrix_eq}
    i\hbar\dot{M}^{(\psi)}=H^{\textnormal{hop}}_{\uparrow}M^{(\psi)} + M^{(\psi)}H^{\textnormal{hop}}_{\downarrow}+V\circ M^{(\psi)}.
\end{equation}
Based on the Lie–Trotter–Suzuki product formula~\cite{trotter:1959,suzuki:1976}, 
we can represent the quantum dynamics through iterative time evolution of matrices:
\begin{equation}
    M^{(\psi)}(t+\delta t)\approx e^{-i\delta t\circ V(t)}\circ e^{-i\delta t H^{\textnormal{hop}}_{\uparrow}}M^{(\psi)}(t)e^{-i\delta t H^{\textnormal{hop}}_{\downarrow}},
\end{equation}
where the symbol ``$\circ$'' denotes the element-wise multiplication and the 
exponential operator in $e^{-i\delta t\circ V(t)}$ means the element-wise 
exponentiation.

\subsection{Full-chain observable} \label{subsec:full_chain}

To illuminate the time evolution process of quantum many-body systems, certain 
experimentally measurable physical quantities are utilized. One such quantity is 
fidelity $\mathcal{F}(t)$, which quantifies the overlap between the initial 
quantum state and its time-evolved counterpart. Mathematically, fidelity is 
expressed as:
\begin{align}\nonumber
    \mathcal{F}(t)=\big|\langle \psi_0 |e^{-i\hat{H}t}|\psi_0\rangle\big|^2
    &=\big|\sum_{\alpha,\beta}M^{*(0)}_{\alpha\beta}M^{(t)}_{\alpha\beta}\big|^2.
\end{align}
When the quantum state extends over the entire chain, the resulting full-chain 
fidelity $\mathcal{F}$ captures the complete quantum information embedded in the 
state.

Another metric is imbalance, which represents the normalized differences in the 
particle occupation numbers between odd and even lattice sites:
\begin{equation}
    \mathcal{I} = \frac{\mathcal{N}_{\textnormal{o}}-\mathcal{N}_{\textnormal{e}}}{\mathcal{N}_{\textnormal{o}}+\mathcal{N}_{\textnormal{e}}},
\end{equation}
where $\mathcal{N}_{\textnormal{o}}$ and $\mathcal{N}_{\textnormal{e}}$ denote the 
occupation numbers at odd and even lattice locations, respectively. The expected 
value of the imbalance for a specific quantum state is calculated as:
\begin{equation}
    \langle \mathcal{I}\rangle = \sum_{\alpha\beta}\left|M_{\alpha\beta}\right|^2\mathcal{I}_{\alpha\beta}.
\end{equation}
The spin-resolved version of the imbalance is defined as:
\begin{equation}
\mathcal{I}^{\uparrow(\downarrow)}=\frac{\mathcal{N}^{\uparrow(\downarrow)}_{\textnormal{o}}-\mathcal{N}^{\uparrow(\downarrow)}_{\textnormal{e}}}{\mathcal{N}^{\uparrow(\downarrow)}_{\textnormal{o}}+\mathcal{N}^{\uparrow(\downarrow)}_{\textnormal{e}}},
\end{equation}
indicating the occupation imbalance for spin-up and spin-down particles. 

\subsection{Accuracy of Trotter decomposition} \label{subsec:accuracy_trotter}

To quantify the numerical errors of simulated observables derived from the Trotter 
decomposition, we use the $L^{p}$-norm~\cite{scherg:2021,yi:2022}:
\begin{align}\nonumber
	|\mathcal{F}_{\textnormal{Runge}}-\mathcal{F}^{n}_{\textnormal{Trotter}}|_{p}&=\left(\int^{\mathcal{T}}_{0}\left|\mathcal{F}_R(t)-\mathcal{F}^{n}_{\textnormal{T}}(t)\right|^{p}dt\right)^{1/p},\\ \nonumber
  |\mathcal{I}_{\textnormal{Runge}}-\mathcal{I}^{n}_{\textnormal{Trotter}}|_{p}&=\left(\int^{\mathcal{T}}_{0}\left|\mathcal{I}_R(t)-\mathcal{I}^{n}_{\textnormal{T}}(t)\right|^{p}dt\right)^{1/p},
\end{align}
with $p=1,2,\ldots$. For $p=\infty$, the norms are defined as:
\begin{align} \nonumber
	|F_{\textnormal{Runge}}-F^{n}_{\textnormal{Trotter}}|_{\infty}&=\textnormal{max}(|F_R(t)-F^{n}_T(t)|),\\ \nonumber
	|\mathcal{I}_{\textnormal{Runge}}|-\mathcal{I}^{n}_{\textnormal{Trotter}}|_{\infty}&=\textnormal{max}(|\mathcal{I}_R(t)-\mathcal{I}^{n}_T(t)|),
\end{align}
where $\mathcal{F}_{\textnormal{Runge}}$ and $\mathcal{I}_{\textnormal{Runge}}$ are 
obtained by the fourth-order Runge-Kutta method with the time step $dt=10^{-4}\tau$, 
equivalent to $n=10^4$ steps per time unit $\tau$. 

\begin{figure*} [ht!]
\centering
\includegraphics[width=\linewidth]{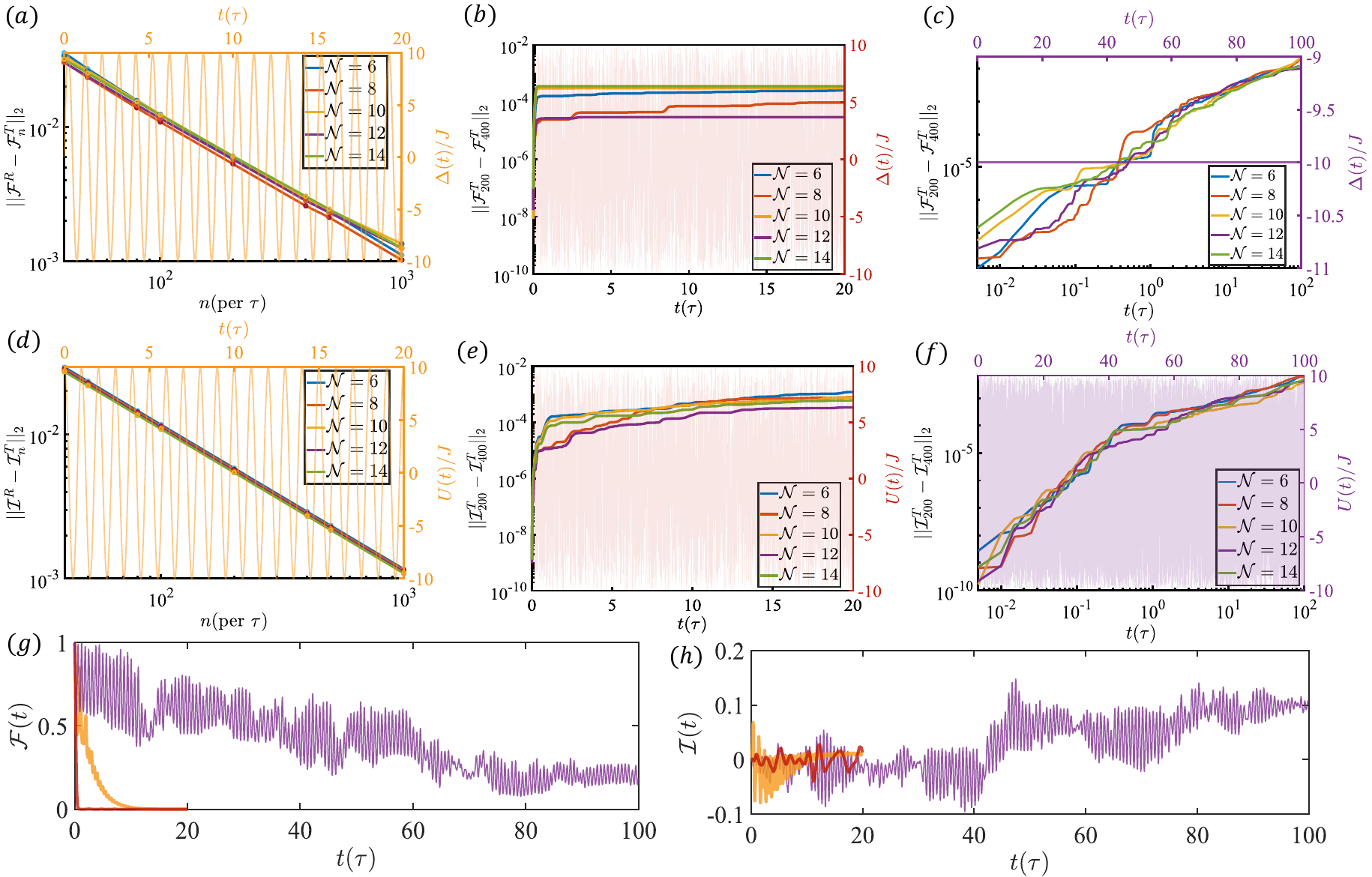}
\caption{Numerical error analysis of Trotter decomposition for the full-chain 
fidelity and imbalance over short- and long-time horizons with various lattice 
numbers. (a,d) Benchmark for Trotter decomposition in the short-time regime 
($\mathcal{T}=20\tau$) obtained by the fourth-order Runge-Kutta method with the 
time step $dt=10^{-4}\tau$. This setup discretizes time into $n=10^{4}$ steps per 
time unit $\tau$. The $L^2$-norm compares the Trotter and Runge-Kutta methods in 
terms of the fidelity and imbalance. The system dynamics are periodically driven 
by $\Delta(t)/J=10\sin(2\pi t)$ and $U(t)/J=10\cos(2\pi t)$, depicted by the orange 
curves and axes. (b,e) $L^2$-norm of the Trotter decomposition errors between 
$n=200$ and $n=400$ steps per unit $\tau$ in the short time regime. The driving 
signals, $\Delta(t)$ and $U(t)$, are uniformly and randomly distributed within the 
range $[-10, 10]J$ over time, represented by the red curves and axes. (c,f) Numerical
error by the Trotter method over a longer time scale ($\mathcal{T}=100\tau$) for 
constant $\Delta=-10J$ and randomly driven $U(t)$, uniformly distributed within 
$[-10, 10]J$, shown as the purple curves and axes. (g,h) The time evolution of 
the full-chain fidelity and imbalance with $n=200$ discrete steps per time unit 
$\tau$ in a system of size $\mathcal{N}=14$.} 
\label{fig:accuracy_trotter}
\end{figure*}

\begin{figure*} [ht!]
\centering
\includegraphics[width=\linewidth]{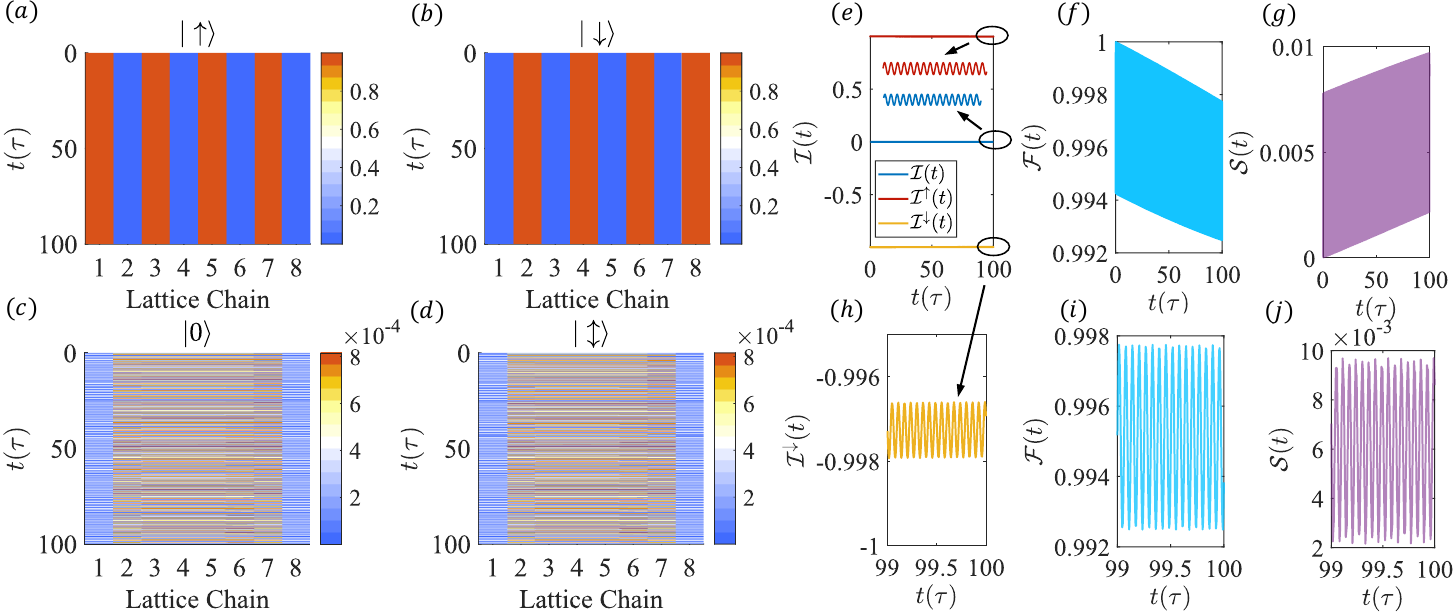}
\caption{Near-ideal Wannier-Stark localization. (a-d) The probability of observing 
the quantum many-body state in four distinct states $|\uparrow\rangle$, 
$|\downarrow\rangle$, $|0\rangle$, and $|\updownarrow\rangle$, at each lattice site. 
(e-j) Time evolution of the nonergodic metrics, including imbalance, full-chain 
fidelity, and half-chain entropy. The initial state is configured as 
$|\uparrow\downarrow\uparrow\downarrow\uparrow\downarrow\uparrow\downarrow\rangle$, 
with a significantly tilted potential $\Delta=100J$ and zero on-site Coulomb 
interaction, $U=0J$, on an $\mathcal{N}=8$ lattice.} 
\label{fig:WS_benchmark_initial_up_down_period}
\end{figure*}

\subsection{Quantum dynamics} \label{subsec:quantum_dynamics}

The Runge-Kutta method, which directly solves the matrix equation~\eqref{eq:matrix_eq},
serves as a benchmark for assessing the Trotter method. Using two different 
discrete steps $n=200$ and $n=400$ per time unit $\tau$ can yield the numerical 
error in the Trotter decomposition. We assess the numerical errors across three 
distinct driven protocols: (1) periodic driving: $\Delta(t)/J=10\sin(2\pi t)$ and 
$U(t)/J=10\cos(2\pi t)$, (2) random driving: $\Delta(t)$ and $U(t)$ uniformly 
distributed in $\in[-10,10]J$, and (3) special protocol: constant $\Delta=-10J$ 
with uniformly random driving $U(t)$ in the interval $[-10,10]J$. For these 
scenarios, the $L^2$-norm metrics of the fidelity and imbalance reveal certain
scaling behaviors across various lattice sizes $\mathcal{N}=6,8,10,12,14$,
as shown in Figs.~\ref{fig:accuracy_trotter}(a-f). These metrics are consistent 
with respect to both the number of discrete steps $n$ per time unit $\tau$ and the 
scaled time $t(\tau)$. It can be seen that the protocol combining constant $\Delta$ 
with random $U(t)$ outperforms the other two control protocols in retaining the 
memory of the quantum state, as illustrated in Fig.~\ref{fig:accuracy_trotter}(g).

\subsection{Sub-chain observable}\label{subsec:sub_chain}

To calculate the observables of a quantum state on a sub-chain, the bases of the 
Hilbert space need to be reorganized and divided into two subspaces: $l$ and $r$. 
Observables in the subspace $l$ are obtained by tracing out the opposing subspace 
$r$. The number of lattice sites in the sub-chain $l$, denoted as $\mathcal{N}_{l}$ and 
counted from the left, defines the scope of this subspace. Accordingly, the number 
of bases in the left-hand $l$ and right-hand $r$ sub-chains are 
$d_{l}=4^{\mathcal{N}_{l}}$ and $d_{r}=4^{\mathcal{N}_r}$ respectively, with the 
relationship $\mathcal{N}_{l}+\mathcal{N}_{r}=\mathcal{N}$. Each lattice site hosts 
one of the four possible states: empty $|0\rangle$, spin up $|\uparrow\rangle$, 
spin down $|\downarrow\rangle$, and doublon $|\updownarrow\rangle$ (spin up and 
down simultaneously). The total number of bases is given by
\begin{equation}\label{eq:basis_entropy}
    d_{l}\times d_{r}=4^{\mathcal{N}},
\end{equation}
reflecting the exponential scaling with the number $\mathcal{N}$ of lattice sites. 
In this framework, the quantum many-body state can be expressed as
\begin{equation}\label{eq:state_matrix_entropy}
    |\psi\rangle = \sum_{l,r}\psi_{l,r}|l\rangle\otimes|r\rangle,
\end{equation}
where $l$ and $r$ are tuples with
\begin{align}\nonumber
    l&\equiv (l_1,l_2,\ldots,l_{\mathcal{N}_l}),\\\nonumber
    r&\equiv (r_1,r_2,\ldots,r_{\mathcal{N}_r}).
\end{align}
Here, $l_i,r_j=0,1,2,3$ corresponds to four possible states for each site. In the
quantum many-body system, particle number is conserved, so the size of the Hilbert 
space for the configuration with $\mathcal{N}/2$ spins up and down is described by 
Eq.~\eqref{eq:basis}. Mapping the focused quantum state into this sub-chain subspace 
without particle number conservation, the elements in the density matrix are sparsely 
distributed and structured as follows:
\begin{align}\nonumber
    \rho = |\psi\rangle\langle\psi|=\sum_{l,l',r,r'}\psi_{lr}\psi^{*}_{l'r'}|l\rangle\langle l'|\otimes|r\rangle\langle r'|.
\end{align}
After tracing out the right-hand sub-chain subspace $r$, the reduced density matrix 
$\rho_l$ is given by
\begin{align}\nonumber
     \rho_l &\equiv \textnormal{Tr}_{r}\rho = \sum_{r''}\langle r''|\psi\rangle\langle\psi|r''\rangle\\\nonumber
     &=\sum_{l,l'}\sum_{r,r',r''}\psi_{lr}\psi^{*}_{l'r'}\delta_{rr''}\delta_{r'r''}|l\rangle\langle l'|\\
    &=\psi\psi^{\dagger}.
\end{align}
Similarly, the reduced density matrix for subspace $r$, denoted as $\rho_r$, can be
obtained as $\rho_r=(\psi^{\dagger}\psi)^{T}$. In Eq.~\eqref{eq:state_matrix_entropy},
the matrix $\psi$ is used to represent the pure quantum many-body state, with the
abstract notations $l$ and $r$ corresponding to the row and column in state matrices, 
respectively.

To enable a calculation of the Von Neumann entanglement entropy in a numerically 
efficient manner, we carry out a singular value decomposition (SVD) of the quantum 
many-body state matrix $\psi$:
\begin{equation}
    \psi_{d_l\times d_r}=U\Sigma V^{\dagger},
\end{equation}
where $\Sigma_{d_l\times d_r}$ is the rectangular diagonal matrix, $U_{d_l\times d_l}$
and $V_{d_r\times d_r}$ are unitary matrices. The nonzero entries along the main 
diagonal of $\Sigma$ represent the real singular values of the matrix $\psi$. 
Consequently, the reduced density matrices can be written rewritten as
\begin{align} \nonumber
	\rho_l &=U\Sigma\Sigma^{\dagger}U^{\dagger}, \\ \nonumber
	\rho^{\textnormal{T}}_r &=V\Sigma^{\dagger}\Sigma V^{\dagger}. 
\end{align}
To calculate the half-chain entropy, we have $d_l=d_r=d$ and 
$\Sigma\Sigma^{\dagger}=\Sigma^{\dagger}\Sigma=\Sigma^2$. The diagonal elements 
$\Sigma^2_i$ of the square matrix $\Sigma^2$ correspond exactly to the eigenvalues 
of the reduced density matrices $\rho_l$ and $\rho_r$. The Von Neuman entanglement 
entropy within the half chain can then be determined by
\begin{align}\nonumber
    S_l = S_r = -\sum_{i=1}^{d}\Sigma^2_{i}\ln\Sigma^{2}_{i}.
\end{align}
Given that $\Sigma^2_i$ can also be interpreted as the square of the singular value 
$\Sigma_i$ of the pure quantum many-body state matrix $\psi$, calculating the 
entanglement entropy primarily involves determining the singular values of this 
state matrix, which has the dimensions $d \times d$.

Another key metric for sub-chains is the fidelity $\mathcal{F}_{\textnormal{sub}}$, 
defined as the overlap between the reduced density matrices $\rho_l$ at the initial 
time $\rho_l(0)$ and at a later time $\rho_l(t)$. Mathematically, 
$\mathcal{F}_{\textnormal{sub}}$ is expressed as
\begin{equation}
    \mathcal{F}_{\textnormal{sub}}(t)=\left(\textnormal{Tr}\left[\sqrt{\sqrt{\rho_l(t)}\rho_l(0)\sqrt{\rho_l(t)}}\right]\right)^{2},
\end{equation}
where the square root of $\rho_l$ is computed as:
\begin{equation}
    \sqrt{\rho_l} = U\Sigma U^{\dagger},
\end{equation}
where the following has been used:
\begin{align}\nonumber
    \sqrt{\rho_l} \sqrt{\rho_l} &= U\Sigma U^{\dagger}U\Sigma U^{\dagger} \\\nonumber
   &=U\Sigma^2 U^{\dagger}\\\nonumber
   &=\rho_l.
\end{align}
The fidelity metric $\mathcal{F}_{\textnormal{sub}}(t)$ thus quantifies how much 
the sub-chain quantum state at time $t$ retains the characteristics of the sub-chain 
state at time $t=0$, according to the evolution of its reduced density matrix.

\subsection{Near-ideal Wannier-Stark localization}\label{subsec:near_ideal_WS}

Ideal Wannier-Stark localization is typically characterized by the freezing and 
localized patterns of particles within the spatial lattice space, often due to the 
influence of a large, tilted electric potential. 
Figures~\ref{fig:WS_benchmark_initial_up_down_period}(a-d) illustrate successful 
particle freezing around the initial state over a long time horizon: 
$\mathcal{T}=100\tau$, under a significantly tilted potential $\Delta=100J$. The 
subsequent time-evolved series of nonergodic metrics closely approximates the ideal 
Wannier-Stark case: 
$\mathcal{I}(t)=\mathcal{I}(0)$, $\mathcal{F}(t)=\mathcal{F}(0)=1$, and 
$\mathcal{S}(t)=\mathcal{S}(0)=0$, in spite of the noticeable decay due to the 
finite tilt of the potential in numerical computations. Effectively,
Fig.~\ref{fig:WS_benchmark_initial_up_down_period} displays near-ideal 
Wannier-Stark localization. Note that Bloch oscillations are still observable in 
the zoom-in region of Figs.~\ref{fig:WS_benchmark_initial_up_down_period}(e,h-j).

\section{Deep reinforcement learning algorithms} \label{appendix_B}

\subsection{PPO pseudo-algorithm}

PPO is a standard, policy-gradient based RL method. It works by alternating between 
gathering data through interactions with the environment and optimizing a surrogate 
objective function that has been clipped. PPO aims to balance exploration and 
exploitation by restricting the magnitude of policy updates. This feature contributes 
to PPO's robustness and efficiency, making it well-suited for a wide range of learning 
tasks. Here we explain the steps and the mathematics behind of the PPO algorithm. 

\vspace*{0.1in}
\noindent\begin{tabular}{p{\columnwidth}}
\hline
\textbf{Pseudo-algorithm:} PPO agent \\
\hline
\textbf{Input:} Initial actor $\pi(a|s;\theta)$, critic $V(s;\phi)$, clipping factor 
$\epsilon$, policy learning rate $\alpha_{\theta}$, value function learning rate 
$\alpha_{\phi}$, number of episodes $N_{\textnormal{ep}}$, number of epochs $K$, 
and number of mini-batches $M$. \\
\textbf{Output:} Optimized policy parameters $\theta$. \\
1: \textbf{for} $Episode=1$ to $N_{\textnormal{ep}}$ \textbf{do}: \\
\quad a: Collect trajectory $\mathcal{D}$ with policy $\pi_{\theta}$. \\
\quad b: Compute advantage estimates $\hat{A}_t$ with $V_{\phi}$. \\
\quad c: Compute return $\hat{\mathcal{G}}_t$. \\
\quad d: Update policy $\pi_{\theta}$ by stochastic gradient ascent and value function $V_{\phi}$ by stochastic gradient descent: \\
\quad\quad \textbf{for} $epoch = 1$ to $K$ \textbf{do}: \\
\quad\quad\quad i: Divide $\mathcal{D}$ into $M$ mini-batches. \\
\quad\quad\quad ii: \textbf{for} each $miniBatch$ in $\mathcal{D}$ \textbf{do}: \\
\quad\quad\quad\quad - Compute probability ratio $r_t(\theta)$. \\
\quad\quad\quad\quad - Compute objective $L^{CLIP}(\theta)$. \\
\quad\quad\quad\quad - Compute square-error loss $L^{VF}(\phi)$. \\
\quad\quad\quad\quad - Update $\theta \leftarrow \theta + \alpha_{\theta} \nabla_{\theta} L^{CLIP}(\theta)$. \\
\quad\quad\quad\quad - Update $\phi \leftarrow \phi - \alpha_{\phi} \nabla_{\phi}L^{VF}(\phi)$. \\
2: \textbf{return} $\theta$. \\
\hline
\end{tabular}
\vspace*{0.1in}

\paragraph*{Generalized advantage estimation.} 
A trajectory is denoted by $\mathcal{D}$:
\begin{align}
    \mathcal{D}=(s_0, a_0, R_0, s_1, \ldots , s_{\mathcal{T}-1}, a_{\mathcal{T}-1}, R_{\mathcal{T}-1}),
\end{align}
which consists of tuples (state $s_t$, action $a_t$, reward $R_t$). We employ 
generalized advantage estimation (GAE)~\cite{schulman:2015}, which leverages a value 
function estimator to calculate the advantage estimates, $\hat{A}_t$, for each time 
step within a trajectory. Specifically, the advantage estimate at time $t$ is given by:
\begin{align}
    \hat{A}_t = \delta_t + (\gamma\lambda)\delta_{t+1} + \ldots +(\gamma\lambda)^{\mathcal{T}-t-1}\delta_{\mathcal{T}-1},
\end{align}
where the temporal difference error, $\delta_t$, is defined as:
\begin{align}
    \delta_t = R_t + \gamma V(s_{t+1};\phi) - V(s_t;\phi)
\end{align}
and $\delta_t$ signifies the immediate advantage of selecting an action under the 
policy $\pi(a_t|s_t;\theta)$. The stochastic policy $\pi(a_t|s_t;\theta)$ represents 
the conditional probability distribution over the action space $a_t$ given the state 
$s_t$. The value function $V(s_t;\phi)$ is employed to evaluate the quality of state 
$s_t$ based on the cumulative reward received. The discount factor $\gamma\in(0,1)$ 
(typical value $\gamma = 0.997$) and the hyperparameter $\lambda$ (commonly 
$\lambda=0.95$) modulate the weighting of future rewards. In essence, the generalized 
advantage $\hat{A}_t$ at time $t$ aggregates discounted future advantages up to the 
terminal stage $\mathcal{T}-1$, enabling more stable and efficient policy updates.

\paragraph*{Return.} The return $\hat{\mathcal{G}}(\mathcal{D})$ is defined as the 
cumulative reward collected throughout a trajectory $\mathcal{D}$, represented by 
$\hat{\mathcal{G}}(\mathcal{D})=\sum_{t=0}^{\mathcal{T}-1}R_t$ with $\mathcal{T}$ 
denoting the time horizon. For ease of mathematical treatment, a discounted version 
is often employed, termed the finite-horizon discounted return: 
\begin{align} \nonumber
\hat{\mathcal{G}}(\mathcal{D})=\sum_{t=0}^{\mathcal{T}-1}\gamma^{t}R_{t}. 
\end{align}
This formulation acknowledges the contribution of future rewards while assigning them 
diminishing importance relative to more immediate rewards. The return at each individual 
time step, $\hat{\mathcal{G}}_t$, is calculated as the sum of the rewards from the 
current time step $t$ onwards, discounted by $\gamma$ to reflect the time value of 
the rewards: 
\begin{align}\nonumber
\hat{\mathcal{G}}_t=\sum_{k=t}^{\mathcal{T}-1}\gamma^{k-t}R_{k}.
\end{align}
Within the proximal policy optimization framework, this return can be derived from the 
generalized advantage estimate by
\begin{align}
\hat{\mathcal{G}}_t = \hat{A}_t + V(s_t;\phi),
\end{align}
where $\hat{A}_t$ represents the advantage estimate at time $t$ and $V(s_t;\phi)$ is 
the value function's estimate of the state's value. 

\paragraph*{Square-error loss and clipped surrogate objective function.} The 
square-error loss, denoted by $L^{VF}(\phi)$, measures how far the value function's 
predictions ($\hat{V}(s_t;\phi)$) are from the actual returns ($\hat{\mathcal{G}}_t$) 
received, which is given by
\begin{equation}
    L^{VF}(\phi) =\hat{\mathbb{E}}_t\left[\left(\hat{ V}(s_t;\phi) - \hat{\mathcal{G}}_t\right)^2\right],
\end{equation}
where $\hat{\mathbb{E}}_{t}[\;\;]$ is the empirical average over a mini-batch of data. 
The clipped surrogate objective function in PPO can be expressed as
\begin{align}
L^{CLIP}(\theta)=\hat{\mathbb{E}}_t\left[\textnormal{min}(r_t(\theta)\hat{A}_t,\textnormal{clip}(r_t(\theta),[1-\epsilon,1+\epsilon])\hat{A}_t)\right],
\end{align}
which ensure that updates to the policy (how the agent decides to act) are not too 
drastic. This is accomplished by using a ``clip'' mechanism that keeps the ratio of 
the new policy to the old policy ($r_t(\theta)$) within a certain range, defined as 
\begin{equation}
    r_t(\theta) = \frac{\pi_{\theta}(a_t|s_t)}{\pi_{\theta_{\textnormal{old}}}(a_t|s_t)}.
\end{equation}
If the new policy is exactly the same as the old one, the ratio is one; Otherwise
the ratio will deviate from one. The clipping keeps this ratio from going beyond the 
specified range, $[1-\epsilon,1+\epsilon]$, which helps slow down policy updates and 
makes learning more stable. 

\begin{figure*} [ht!]
\centering
\includegraphics[width=\linewidth]{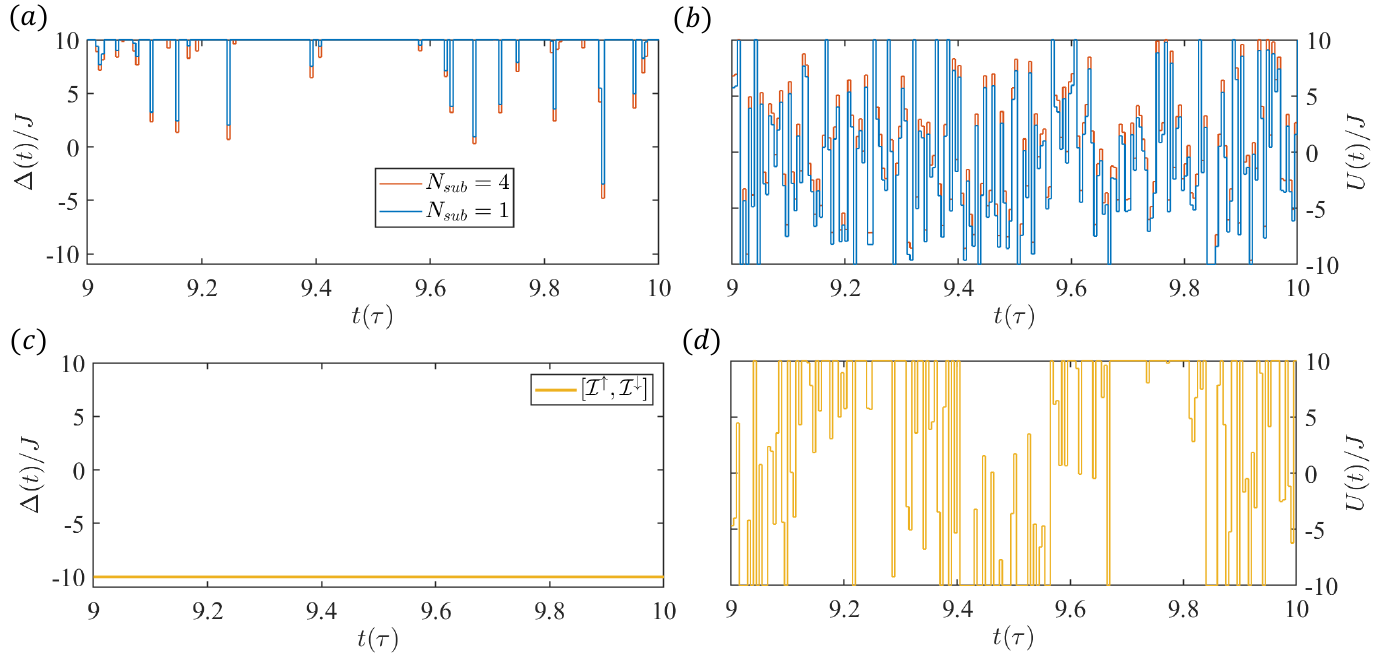}
\caption{The zoomed-in view of the optimized nonergodic control flow from $9\tau$ to $10\tau$  is shown for Figs.~\ref{fig:numLattice_14_T_5_F_sub_imbalance}(e) and~\ref{fig:numLattice_14_T_5_F_sub_imbalance}(f). Panels (a) and (b) display the optimized control flow when the observation is the fidelity with respect to $N_{sub}=4$ and $N_{sub}=1$, respectively. Panels (c) and (d) show the optimized control flow when the observation is the imbalance  $[\mathcal{I}^{\uparrow}, \mathcal{I}^{\downarrow}]$.}
\label{fig:control_zoom_in}
\end{figure*}

\section{Extended discussion on optimized control flow}

The appearance of whether the optimal control flow is ``wild" depends on the time scale. Since our time step size $dt$ is constant, the control flow is optimized as piecewise constant control signals, as illustrated in the zoomed-in view in Fig.~\ref{fig:control_zoom_in}. Recent work~\cite{werninghaus:2021} and related theoretical studies~\cite{motzoi:2009} demonstrate that generating square waves on a $1$ ns time scale is experimentally feasible in superconducting platforms, which are viable candidates for simulating the 1D tilted Fermi-Hubbard model~\cite{guo:2021,reiner:2016,zhang:2023}. In our manuscript, the time step is set to $dt=0.005 \tau$, where $\tau$ is determined by the nearest-neighbor coupling $J$ as $\tau=\hbar/J$. Consequently, a $1$ ns time step corresponds to a coupling strength $J$ on the order of $1$ MHz, which is realistically achievable in superconducting qubits~\cite{barends:2014, chen:2014}. A reward function similar to the one used in our work has been applied to the single-particle quantum control in the upside-down potential~\cite{wang:2020cartpole} and the nonlinear double-well potential~\cite{borah:2021}, demonstrating the versatility and effectiveness of our optimized reward function. Additionally, we explored incorporating the control field into the reward function, but this resulted in significantly worse performance. 

The fast Fourier transformation (FFT) analysis indicates that the solutions for $\Delta(t)$ and $U(t)$ from Figs.~\ref{fig:numLattice_14_T_5_F_sub_imbalance}(e) and~\ref{fig:numLattice_14_T_5_F_sub_imbalance}(f) do not have dominant Fourier modes but instead rely on a broad spectrum of frequencies. This complex frequency structure is consistent with the requirements of optimal nonergodic control, which involves intricate manipulation of the system dynamics. Based on this, we developed a DRL-aligned protocol in a Sec. IV for the physical interpretation and insights.

\bibliography{QMBS_RL}

\end{document}